\documentclass[onecolumn]{webofc}
\usepackage[varg]{txfonts} 
\usepackage[normalem]{ulem}
\usepackage{amsfonts}
\usepackage{subcaption}
\usepackage{dsfont}
\usepackage{graphicx}
\usepackage{dcolumn}
\usepackage{bm}
\usepackage{physics}
\usepackage{xcolor}
\usepackage{mathrsfs}
\usepackage{mathtools}
\usepackage{comment}

\usepackage{orcidlink}
\begin{document}
\title{\textit{Cahier de l'Institut Pascal}\\ Noisy Quantum Dynamics and Measurement-Induced Phase Transitions }

\author{\firstname{Alexios} \lastname{Christopoulos \orcidlink{0000-0003-4857-5922}}\inst{1} \and
\firstname{Alessandro} \lastname{Santini 
\orcidlink{0000-0002-4949-7463}}\inst{2}
        \and
        \firstname{Guido} \lastname{Giachetti \orcidlink{0000-0002-4928-7693}}\inst{1} 
}

\institute{
           Laboratoire de Physique Theorique et Modelisation, CY Cergy Paris Universite, CNRS, F-95302 Cergy-Pontoise, France \and SISSA, via Bonomea 265, 34136 Trieste, Italy 
          }
\pagestyle{plain}
\abstract{This is a conference proceeding in the framework of workshop “OpenQMBP2023”\footnote{\href{https://indico.ijclab.in2p3.fr/event/8925/}{OpenQMBP2023}: New perspectives in the out-of-equilibrium dynamics of open many-body quantum systems
Jun 12–30, 2023 at Institut Pascal (Orsay, France).} at Institute Pascal (Orsay, France) and associated to the lecture given by Prof. Ehud Altman. We provide a comprehensive analysis of recent results in the context of measurement-induced phase transitions (MIPT) in quantum systems, with a particular focus on hybrid quantum circuits as a model system in one-dimension. Recent resutls, demonstrate how varying the rate of projective measurements can induce phase transitions, resulting in abrupt changes in the properties of the entanglement. The interplay between unitary evolution and measurement processes  can be investigated, through mappings to classical statistical models and the application of replica field theory techniques. 
Starting from a low-entangled state, there can be three regimes characterized by different dynamics of bipartite entanglement entropies for a portion of the system: high-rate measurements leading to rapid entanglement saturation (area law), low-rate measurements allowing linear entanglement growth (up to volume law), and a critical rate at which entanglement grows logarithmically. Finally, we present results on the non-local effects of local measurements by examining the field theory of critical ground states in Tomonaga-Luttinger liquids. }
\maketitle
\section{Introduction}
To begin with, it is important to mention that the goal of this proceeding is to present a compact inspection of recent results on MIPTs and not an extensive review of the field. For more details, the reader can refer to the bibliography presented below as well as to other  reviews on the field such as \cite{Potter2022,Lunt2022}.

In the field of quantum mechanics, measurements assume a central role. Beyond being useful for observing and characterising quantum systems, their back-action affects as well the dynamics of a quantum system.

In particular, measurements can disrupt the quantum correlations between different parts of the system, inhibiting quantum evolution, as seen in phenomena such as the quantum Zeno effect \cite{Degasperis1974DoesTL,PhysRevResearch.2.033512,Biella2021manybodyquantumzeno,lami2023continuously}. However, sparse measurements allow for the formation of complex quantum states, characterised by an extensive scaling of their entanglement entropy with the system's volume. More generally, the competition between the unitary dynamics, which tends to establish quantum correlations, and the measurement process can give rise to measurement-induced phase transitions (MIPT), visible, e.g., by looking at the bipartite entanglement. These phenomena have been extensively investigated across various domains, including quantum circuits \cite{qcircuits_1,qcircuits_2,Fisher2018PRB,qcircuits_4,Nahum2019PRX,Altman2020PRB,qcircuits_6}, fermionic models \cite{DeLuca2019SciPost,PhysRevB.105.094303,PhysRevB.108.104313,lumia2023measurement} and quantum spin systems \cite{10.21468/SciPostPhys.15.3.096,PhysRevB.105.064305}, quantum hardware \cite{Hoke2023MeasurementinducedEA}, trapped ions \cite{ions1,ions2,ions3} and trapped atoms \cite{ions1}, with a broad array of tools that spans quadratic fermionic models, tensor-networks, replica symmetry breaking \cite{fava2023nonlinear,giachetti2023elusive,popperl2023localization}, and mappings onto toy models \cite{Nahum2019PRX,ha2024measurementinduced,PhysRevB.109.125148}.

In this proceeding, we present shorty the findings on the interplay between a noisy unitary evolution and repeated measurements of local degrees of freedom, which are generally indicated as \textit{the monitoring process}. This implies the presence of time-dependent randomness both in the non-Hermitian measurement scheme and in the hermitian unitary evolution. This setup enables the derivation of analytical results in the context of MIPT through various mappings of the quantum problem onto well-established classical statistical toy models. Finally, we will analyse some recent findings on how partial local measurements disrupt the scaling of entanglement of critical ground states from a different viewpoint.

The proceeding is organised as follows. In Sec. \ref{sec:Measurements_in_QM}, we highlight recent results on some theoretical aspects of measurements in quantum mechanics, followed by a brief simple argument that shows non-local effects arising from quantum measurements in Sec. \ref{sec:non-local-effect}. Section \ref{sec:MIPT_in_RUC} presents both numerical and analytical results within the context of monitored random unitary circuits and their mapping to classical statistical mechanics models. Finally, in Sec. \ref{sec:MIPT_in_GS}, we extend our investigation to the recent discoveries of the impact of partial measurements on the critical ground states of quantum systems.

\section{Measurements in quantum Mechanics} \label{sec:Measurements_in_QM}
The standard way to introduce measurement in quantum mechanics is to start with projective measurements\cite{wiseman_milburn_2009}, also commonly known as von Neumann measurements or strong measurements \cite{von1955mathematical}.

Let us consider a physical observable $O$, associated to the hermitian operator, $\hat{O}$ that we want to measure. This operator can be decomposed as $\hat{O} = \sum_\mu o_\mu \hat{P}_\mu$, where {$o_\mu$} are the (real) eigenvalues (which we assume to be a discrete set for the sake of simplicity) and $\hat{P}_{\mu}$ the projector operator on the corresponding eigenspace. Furthermore, in the case of non-degenerate spectrum, the projection operator $\hat{P}_{\mu}$ can be written as $\hat{P}_{\mu}= \dyad{\mu}$, $\ket{\mu}$ being the normalized eigenvector corresponding to the eigenvalue $o_\mu$. Therefore, we have 
\begin{equation}
    \hat{O} = \sum_\mu o_\mu \hat{P}_\mu = \sum_\mu o_\mu \dyad{\mu}.
\end{equation}
If $\hat{\rho}$ is the density matrix of the system, the projective measurement of the observable $\hat{O}$ returns one of the eigenvalues $o_\mu$ with a probability $\mathscr{P}_\mu = \Tr{\hat{\rho} \hat{P}_\mu}$. As a consequence of the back action of the process on the system, after the process the state  of the system is projected on the eigenspace corresponding to $o_\mu$, i.e.
\begin{equation}
\label{eq:projrho}
    \hat{\rho}_\mu = \frac{\hat{P}_\mu \hat{\rho} \hat{P}_\mu}{\Tr{\hat{\rho} \hat{P}_\mu}} ,
\end{equation}
hence the name projective measurement. 

On the other hand, it is possible to describe more general types of measurements, also called weak measurements, still based on the paradigm of projection in Eq.~\eqref{eq:projrho} but within an enlarged Hilbert space, which includes the system and some modelling of the detector itself. In these cases,  
one obtains less information about the system, while at the same time perturbing it to a lesser extent. For example, in photodetection \cite{RevModPhys.86.1391}, measuring the external field (light) acts as an indirect projective measurement of the quantum system. One  analyses properties of the detected light (frequency, polarization) to gain information about the system's possible energy state transitions. This is analogous to using light's characteristics as clues to infer the state of a system.

The following toy model gives a straightforward example of a weak projective measurement. Let us consider a two-level ancilla, represented by the states $\{\ket{+},\ket{-}\}$ eigenstates of the Pauli matrix $\sigma_z$, initially prepared in the state\begin{equation}
    \ket{\alpha} = \frac{\ket{+}+\ket{-}}{\sqrt{2}},
\end{equation}
coupled to the system of interest, represented by the state $\ket{\psi_t}$. Let both the ancilla and the system evolve over a time $T$ under the unitary evolution operator,  $\hat{U}_{S+A}(T)$ \begin{equation}
    \hat{U}_{S+A}(T) \ket{\psi_{t}}\ket{\alpha} = \left(\hat{K}_{+}\ket{\psi}\right)\ket{+}+\left(\hat{K}_{-}\ket{\psi_t}\right)\ket{-}, 
\end{equation}
where $\hat{K}_{\pm}= \mel{\pm}{\hat{U}_{S+A}(T)}{\alpha}$ act solely onto the Hilbert space of the system. After this evolution, a projective measurement is performed on the ancilla along the $z$-axis, leading to the outcome $\sigma = \pm$. Consequently, the back-action of the measurement bring the system in the state
\begin{equation}
    \ket{\psi_{t+T}} = \frac{\hat{K}_{\sigma}\ket{\psi_t}}{\sqrt{\mel{\psi_t}{\hat{K}_{\sigma}^\dagger \hat{K}_{\sigma}}{\psi_t} }}.
\end{equation}
with $\sigma=\pm$ depending on the outcome of the projective measurement on the ancilla. Equivalently, by working in terms of density operators,
\begin{equation}
    \hat{\rho}_\sigma = \frac{\hat{K}_\sigma \hat{\rho} \hat{K}_\sigma^\dagger}{\Tr{\hat{K}_\sigma \hat{\rho} K^\dagger_\sigma }} \, , 
\end{equation}
while the probability of each outcome is given by 
\begin{equation}
\label{eq:prob}
    \mathscr{P}_\sigma = \Tr{\hat{K}_\sigma \rho {\hat{K}^{\dagger}_\sigma}} \ .
\end{equation}
More generally, any physical evolution of the system induced by unitary dynamics and projective measurements over an enlarged Hilbert space can be encoded in a set of so-called Kraus operators $K_\mu$, $\mu=1, \dots,m$, \cite{Griffiths2010QuantumCK} i.e. a system of operators acting only on the system Hilbert state and such that $\sum^m_{\mu=1} K_\mu^{\dagger} K_\mu = 1$. In the case, in which  $\hat{K}_{\pm}$ are just projectors within the system Hilbert space, the process reduces to the usual projective measurement. This situation can be seen as a limiting case when the system and the ancilla become highly entangled, so that projective measurement of the former acts as a projective measurement of the latter as well. We can thus interpret the coupling between the ancilla and the system as a way of quantifying the measurement strength.  

As measurement is not a deterministic process, the change in the state of the system depends on the result of the measurement.
\begin{figure}
    \centering
\includegraphics[width=0.6\linewidth]{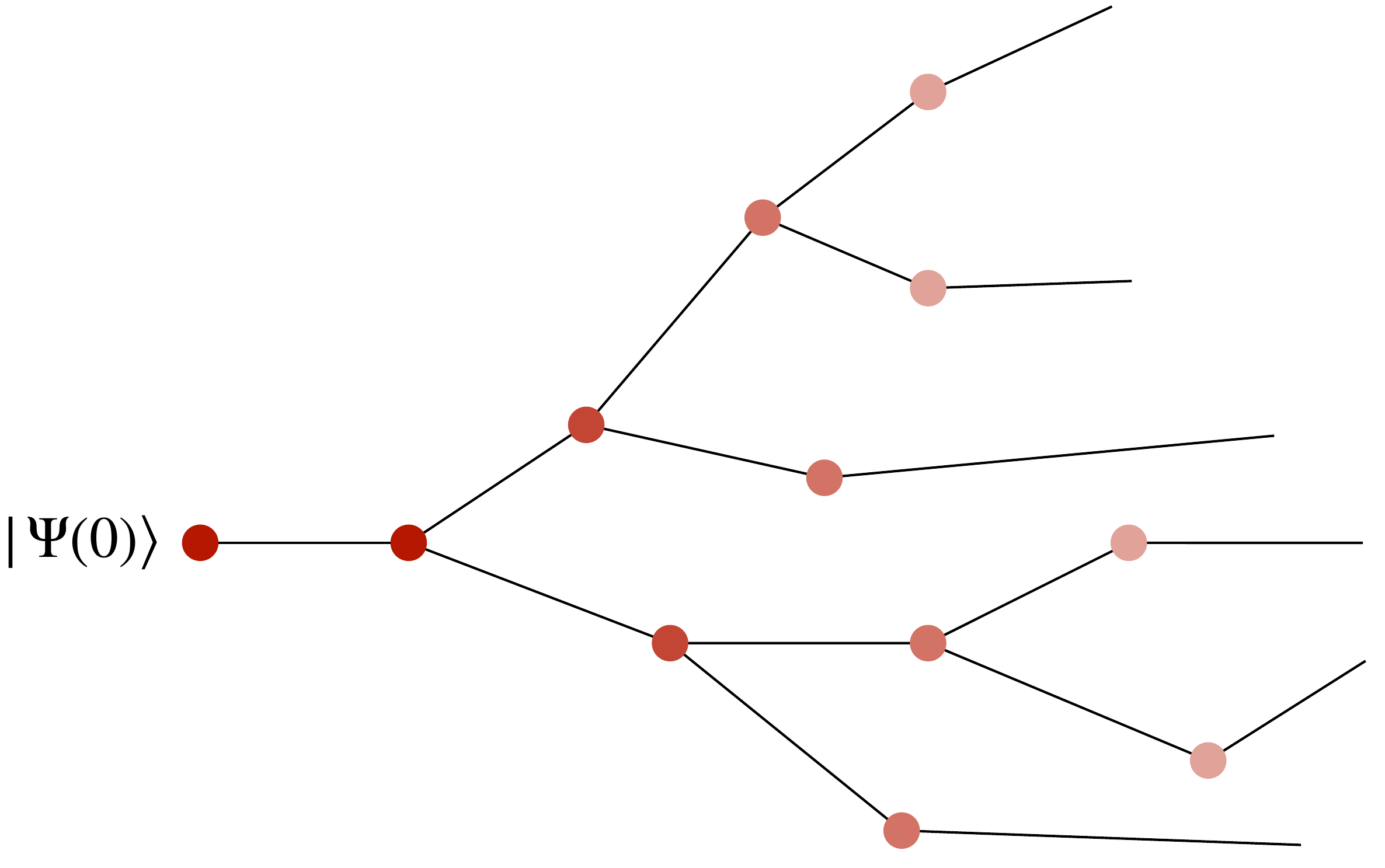}
    \caption{\textbf{Quantum Trajectory:} the dots represent measuring process after which the state of the system change according to the outcome probabilities.}
    \label{fig:QuantumTrajectory}
\end{figure}
Let us now introduce the \emph{unconditional}, i.e. average system's state without any information about the measurement outcomes: 
\begin{equation}
    \hat{\rho}' = \sum_\mu \mathscr{P}_\mu \hat{\rho}_\mu = \sum_\mu K_\mu \hat{\rho} K^\dagger_\mu. \label{eq:unconditional_state} \ . 
\end{equation}
Let us notice that this map preserves the normalization and definite positivity of the density operator. However, the unconditional state is formally equivalent to discarding the measurement results. For this reason, the evolution leads to less and less information in the state and in the absence of additional structure, it eventually drives the system towards an infinite temperature, completely mixed state. 

On the other hand, the \emph{conditional} state of a quantum system refers to its post-measurement state conditioned to specific measurement outcomes. This conditional state precisely reflects the quantum state of the system, informed by the acquired knowledge of these measurement results. When measurements are repeated over time, the sequence of states conditioned on the measurements up to that point is also called a quantum trajectory. Notably, a protocol that measures the system sequentially at different times gives rise to a complex tree of potential quantum trajectories. Specifically, the protocol leads to exponentially many distinct trajectories, corresponding to the possible outcomes from the corresponding series of measurements. This is illustrated in Fig. \ref{fig:QuantumTrajectory}, which provides a qualitative sketch showing how different measurement outcomes can lead to various possible quantum trajectories.

\section{Local Measurements have a non-local effect on entangled states} \label{sec:non-local-effect}

\begin{figure}
    \centering
    \includegraphics[width=0.6\linewidth]{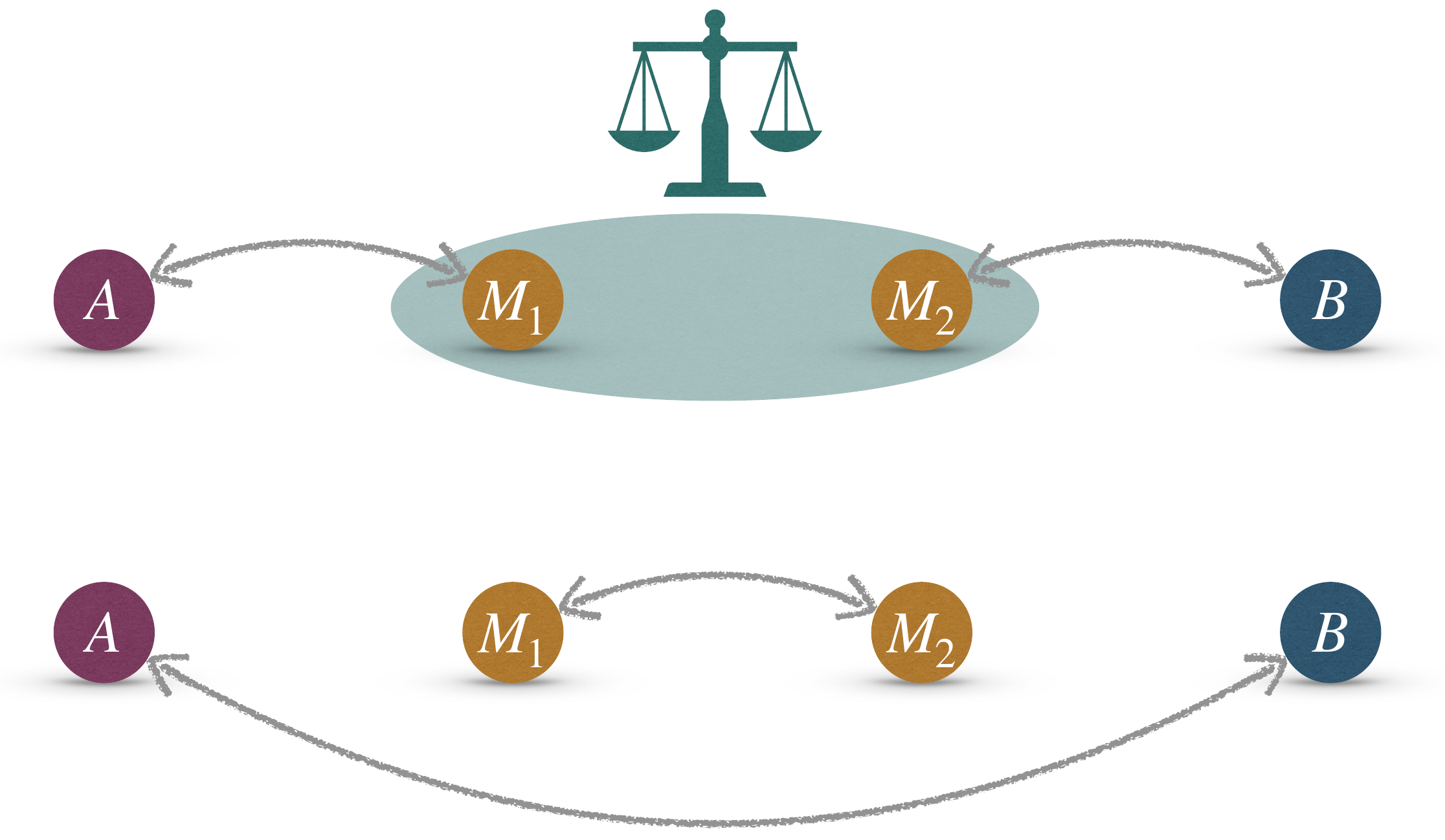}
    \caption{\textbf{Entanglement Swapping}: initially $(A,M_1)$ and $(M_2,B)$ are entangled. After measuring $M=(M_1,M_2)$ in the Bell basis, $(A,B)$ become entangled, establishing a quantum correlation previously absent. }
    \label{fig:AliceBob}
\end{figure}

Before analysing the role that projective measurements play in many-body systems, as a warm-up let us start our analysis by exploring the phenomenon of measurement-induced entanglement swapping, demonstrating the non-local effects that local measurements can have on entangled states; see Fig. \ref{fig:AliceBob} for a scheme of the process. 

To illustrate this concept, we consider a simple four-qubit system involving two distant components, Alice (denoted as qubit $A$) and Bob (denoted as qubit $B$), and a composite system $M$ represented by qubits $M_1$ and $M_2$, which will be measured.

Let us introduce the Bell basis $\{ \ket{\Phi^{\pm}},\ket{\Psi^{\pm}} \}$ of a pair of qbits
\begin{align}
    \ket{\Phi^{\pm}} = \frac{1}{\sqrt{2}}  \left(\ket{00}\pm\ket{11}\right), \\ \ket{\Psi^{\pm}} = \frac{1}{\sqrt{2}}\left(\ket{01}\pm\ket{10}\right).
\end{align}
We begin with an initial state where Alice and Bob share a Bell state with the composite system $M$, so that the global system state $\ket{\psi}$ is given by
\begin{align}
    \ket{\psi} &= \ket{\Phi_{AM_1}^+}\ket{\Phi_{M_2B}^+} \\
    &= \frac{1}{2}\bigl(\ket{0_A0_{M_1}}+\ket{1_A1_{M_1}}\bigr)\bigl(\ket{0_{M_2}0_B}+\ket{1_{M_2}1_B}\bigr) \notag.
\end{align}
Next, we want to perform a projective measurement on the composite system $M$ in the Bell basis. To this extent it is convenient to write the state $\ket{\psi}$ in terms of the Bell pairs of $M$
\begin{align}
  \notag  \ket{\psi} = \frac{1}{2}\biggl[&\ket{0_A0_B}\frac{\ket{\Phi^+_M}+\ket{\Phi^-_M}}{\sqrt{2}}+\ket{1_A1_B}\frac{\ket{\Phi^+_M}-\ket{\Phi^-_M}}{\sqrt{2}} \\+&\ket{0_A1_B}\frac{\ket{\Psi^+_M}+\ket{\Psi^-_M}}{\sqrt{2}}+\ket{1_A0_B}\frac{\ket{\Psi^+_M}-\ket{\Psi^-_M}}{\sqrt{2}} \biggr] \, .
\end{align}
The measurement results in the generation of one of the following entangled pairs between Alice and Bob, each occurring with equal probability
\begin{align}\notag
   \ket{\psi_1} = \ket{\Phi_{AB}^+}\ket{\Phi^+_M}, &&
   \ket{\psi_2} =\ket{\Phi_{AB}^-}\ket{\Phi^-_M},\\
\ket{\psi_3} = \ket{\Psi^+_{AB}}\ket{\Psi^+_M},&& \ket{\psi_4} = \ket{\Psi^-_{AB}}\ket{\Psi^-_M}.
\end{align}
Our simplified model highlights the non-local characteristics of quantum systems, where local measurements performed on one segment of the system can instantaneously influence the entangled relationships among distant particles. Specifically, when measurements are conducted non-locally in the Bell basis---a fundamental set of states used to demonstrate quantum correlations---the system $M$ establishes a quantum correlation between components $A$ and $B$ that was not previously present. This behaviour underscores the intriguing aspect of quantum entanglement, where actions on one part of the system can affect other parts instantaneously, regardless of the distance separating them. This phenomenon, often described as “spooky action at a distance”, vividly illustrates one of the most profound and debated aspects of quantum theory, as discussed in foundational works such as J.S. Bell's paper on the EPR paradox and non-locality \cite{PhysicsPhysiqueFizika.1.195}.

\section{Measurement Induced phase transitions in random unitary circuits} \label{sec:MIPT_in_RUC}

\begin{figure}[t]
    \centering    \includegraphics[width=0.7\linewidth]{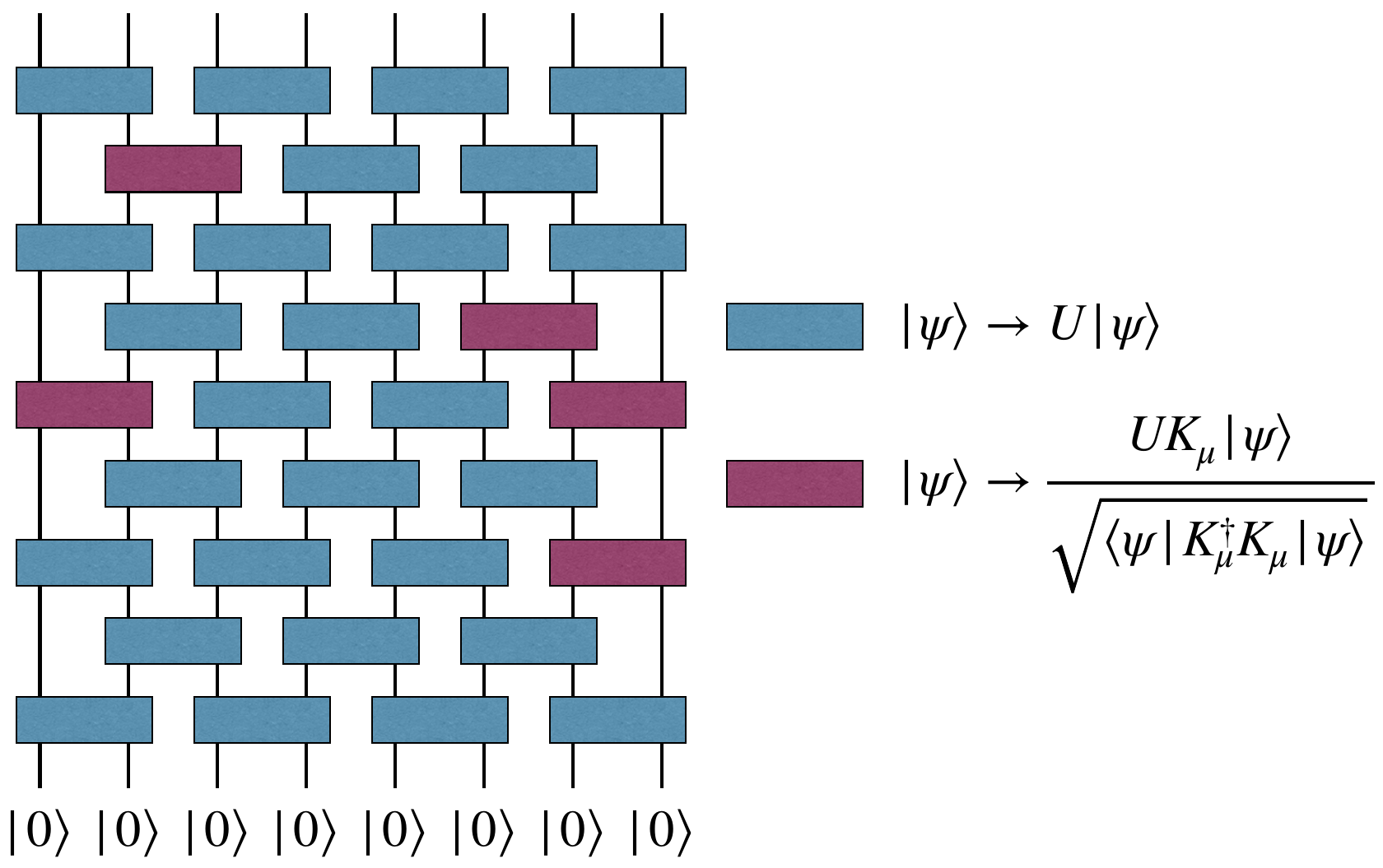}
    \caption{\textbf{Brick wall circuit:} scheme of a quantum trajectory. In blue, we marked the unitaries applied to the qubits, while in purple the measurement operators.}
    \label{fig:brickcircuit}
\end{figure}

This section highlights the competition of measurements and unitary dynamics in the context of many-body systems.
We present recent findings\cite{Fisher2018PRB, Nahum2019PRX, Altman2020PRB} related to the criticality and universal features of MIPT in many-body quantum systems. Our analysis will focus on one-dimensional \textit{hybrid} quantum circuits, which incorporate both unitary gates and projective measurements.

The system under consideration consists of an array of $N$ 2-qubits, initially prepared in a product state.
\begin{equation}
\ket{\Psi_0} = \bigotimes_{i=1}^N \ket{\psi_i}.
\end{equation}
One employs a brick wall circuit approach, as illustrated in Fig.~\ref{fig:brickcircuit}, where random unitary operations $\hat{U}$ are sequentially applied to all the even pairs first, and then to the odd pairs in the subsequent time-step. The sequence of operations alternates between even and odd pairs to simulate the propagation of interactions through the system. Each application of $\hat{U}$ is followed, with a probability $p$, by a quantum measurement. The composite action of the unitary operation followed by the measurement is mathematically represented as $\hat{U}\hat{K}_\mu$, where $\hat{K}_\mu$ denotes the measurement operator. This model captures the dynamics of quantum states as they evolve under random, controlled disturbances introduced by the unitary transformations and the stochastic nature of quantum measurements. Such a setup is used to investigate phenomena like quantum walks, entanglement propagation, or decoherence processes in a controlled quantum system.

A key observable to understand the physics of such models is the averaged entanglement entropy of an interval $A$: given a set of quantum trajectories, each one corresponding to the sequence of measurement outcomes $\vec{\xi} = \{\xi_{1}, \xi_{2}, \xi_{3}, ...\}$, it corresponds to the entanglement entropy of $A$ averaged both over all quantum trajectories and over the random unitaries
by\begin{equation}
    S_A(t) = \overline{\sum_{\vec{\xi}} P_{\vec{\xi}} S(\hat{\rho}_A(t)}\label{eq:average_SRUC} \, .
\end{equation}
Here, we denote with $P_{\vec{\xi}}$ the probability of the quantum trajectory $\vec{\xi}_\mathbf{m}$, while the bar denotes the average over the random unitary gates. When discussing averaged entanglement entropy, one considers multiple possible quantum trajectories that a system can take due to different sequences of measurement outcomes. For each trajectory, one computes the entanglement entropy after each measurement. To obtain the averaged entanglement entropy, one then takes the statistical average of these entanglement entropy values across all the trajectories. By analysing the averaged entanglement entropy, one can infer the overall behaviour of the system under different types of measurements and interactions, assessing the robustness of quantum entanglement against environmental interactions and measurement-induced disturbances \cite{entangS1}, as well as the decoherence \cite{entangS2}.

\subsection{Numerical results}
Preliminary, let us analyse some numerical findings, based on Li et al. \cite{Fisher2018PRB}. In this work, the cases in which random unitaries are drawn from Haar measure or Clifford orbits are considered, while both the cases in which the two-qubit measurements are Bell pairs measurements or local measurements are considered.  Notably, a phase transition was observed in all the settings analysed, challenging the previous expectations that such a transition did not occur, according to Ref.\cite{DeLuca2019SciPost}, in which monitored systems with symmetry $U(1)$ were studied. The absence of phase transition in this latter model could be traced back to the fact that, within non-interacting fermionic systems as the one in \cite{DeLuca2019SciPost}, (local) measurements are efficient in eliminating long-range entanglement from their highly fragile Bell pairs.

At long-times, a stationary value of the entanglement entropy is attained, and it was found to scale as
\begin{equation}
S_A (p,cN) \sim N^\gamma F\left[(p-p_c)N^{1/\nu}\right],
\end{equation}
with $c=\ell_A/N$, $\ell_A$ being the length of the \emph{extensive} interval $A$. 
The exponents $\gamma$ and $\nu$ depend on the specific choice of the evolution protocol, while the scaling function $F(x)$ has the following asymptotic scaling

\begin{equation}
F(x) \sim 
\begin{cases}
    |x|^{(1-\gamma)\nu} & \text{as } x\to -\infty\\
    |x|^{-\gamma\nu} & \text{as } x\to \infty
\end{cases},
\end{equation}
so that
\begin{equation} \label{eq:scalingSnu}
\lim\limits_{{N\to \infty}} S_A (p,cN) \sim\begin{cases}
     N |p-p_c|^{(1-\gamma)\nu} & \text{for } p< p_c\\
      |p-p_c|^{-\gamma\nu} & \text{for } p> p_c
\end{cases}.
\end{equation}
These numerical findings show that the stationary  entanglement entropy  demonstrates a phase transition at some critical measurement probability $p_c$, between volume and area law, so that $S/N$ can effectively serve as an order parameter for this MIPT. 
In particular: \newline
\indent \textbf{Volume-Law Phase}: When the stationary entanglement entropy is proportional to the volume of the interval $\ell_A$ for $p < p_c$, the system exhibits a \emph{volume-law phase} \cite{PRXQuantum.3.030201}. This phase is characterized by strong, extensive quantum correlations that permeate the entire system. Typically observed in less-disrupted quantum states, the volume-law phase suggests a robust entangled state where quantum information is widely distributed throughout the system.

\textbf{Area-Law Phase}: In contrast, when the stationary entanglement entropy remains constant regardless of the size of $\ell_A$ for $p > p_c$, it indicates an \emph{area-law phase} \cite{RevModPhys.82.277}. This phase emerges in highly disordered or measurement-intensive scenarios, leading to decoherence and localization of quantum states. Here, entanglement is primarily confined to the boundaries or interfaces, and does not scale with the size of the subsystem, reflecting localized states with minimal entanglement spread.

\begin{figure}
    \centering    \includegraphics[width=0.6\linewidth]{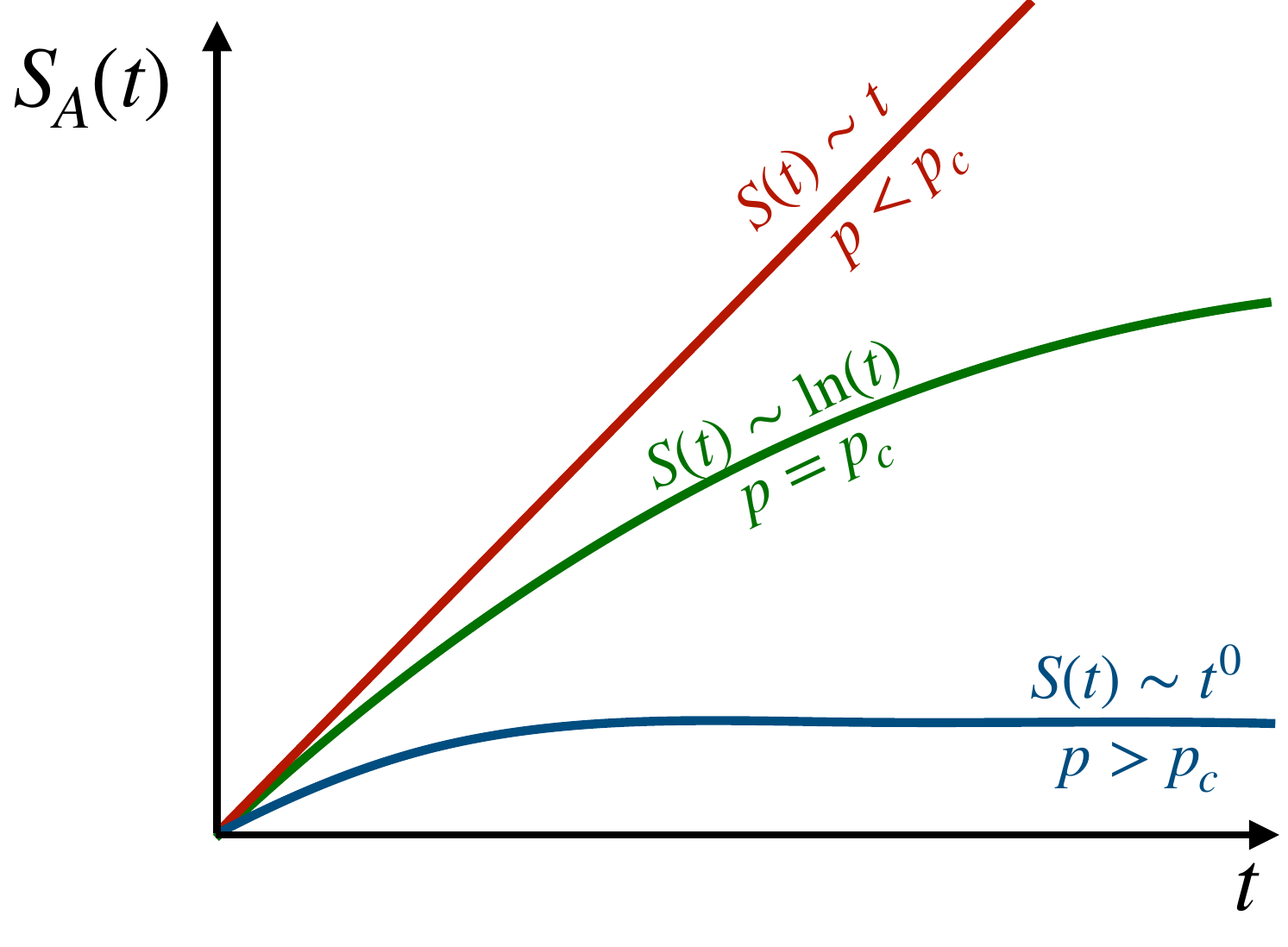}
    \caption{\textbf{Entanglement growth:} varying the probability of applying a measurements' operator $p$ there are three different phases, for $ p  >p_c$ we are in the \textit{Quantum Zeno} phase in which there is no growth of entanglement, if $ p<p_c$ the measurements are too sparse and the entanglement as a volume-law growth, finally for $p=p_c$ we are in the critical phase in which the entanglement grows logarithmically. }\label{fig:entanglemententropy_sketch}
\end{figure}

\subsection{Mapping to an optimisation problem in classical percolation}

\begin{figure}
    \centering
    \includegraphics[width=0.5\linewidth]{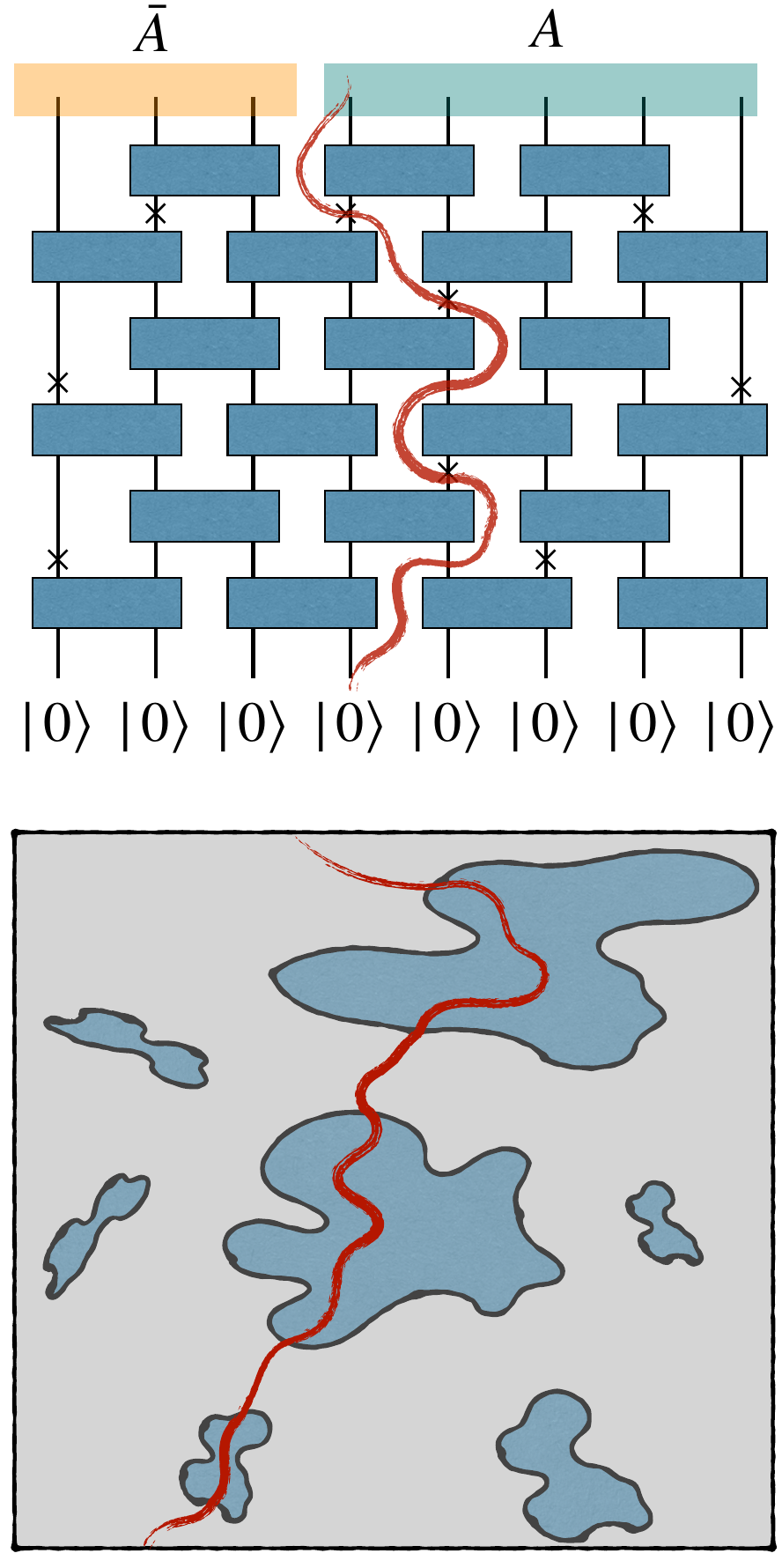 }
    \caption{\textbf{Minimal Cut:} a minimal cut path. Top panel: The rectangles represent random unitaries, while the $\times$ random projective measurements. {\textbf{Bottom panel:}} sketch of the minimal cut traversing  clusters for $p=p_c$. The minimal cut is the path that separates the system into two partitions by passing through the least number of bonds.
}
    \label{fig:Percolation}
\end{figure}
In this section, we consider the random unitary circuit already investigated by Skinner et al. \cite{Nahum2019PRX}, in which the random unitaries $\hat{U}$ are drawn from the Haar measure, while $K_\mu$ are projectors on the computational basis $\{ \ket{0}, \ket{1} \}$. The scaling in time of $S_A(t)$ of an extensive interval can be here understood by introducing a mapping with a solvable statistical toy model; the relevant results being summarized in Fig.\,\ref{fig:entanglemententropy_sketch}. Specifically, in the thermodynamic limit, for $p<p_c$ an asymptotic linear growth of $S_A (t)$ is observed (volume-law phase), while for $p < p_c$ it saturates to a finite value (area-law phase). At the border between the two phases ($p=p_c$) a logarithmic scaling was found.  

The aforementioned mapping can be made exact by considering, instead of the von Neumann entropy, the zeroth order R\'{e}nyi entropy $S_{0,A}$ as a proxy for the qualitative behaviour of more refined entanglement measures. Let us notice that $S_{0,A}$ can be thought as the logarithm of the number $n$ of non-zero eigenvalues of the reduced density matrix. In the absence of projective measurements ($p=0$), it can be shown \cite{qcircuits_4} that in turn $n$ corresponds to the minimum number of bonds in the circuit that must be crossed to isolate the physical “legs” (external bonds at the top boundary) of the qubits within the subsystem $A$ from those outside. As projective measurements locally fix the state of the measured qubit to $\ket{1}$ or $\ket{0}$, they effectively \textit{break a bond} in the network, and the cost of crossing this already broken bond reduces to zero. As a consequence, for $p >0$, for a given realization of the gates, $n$ can be defined as the minimal number of \textit{unbroken} bonds that must be crossed to isolate the physical legs of $A$ (minimal cut). Consequently, the computation of $S_{0,A}$ for a subsystem transforms into an optimization problem within a classical percolation configuration, aiming to determine the minimum number of bonds to cut to isolate the subsystem from the rest of the network, see Fig. \ref{fig:Percolation} for a scheme.  
\begin{figure}
    \centering
    \includegraphics[width=0.5\linewidth]{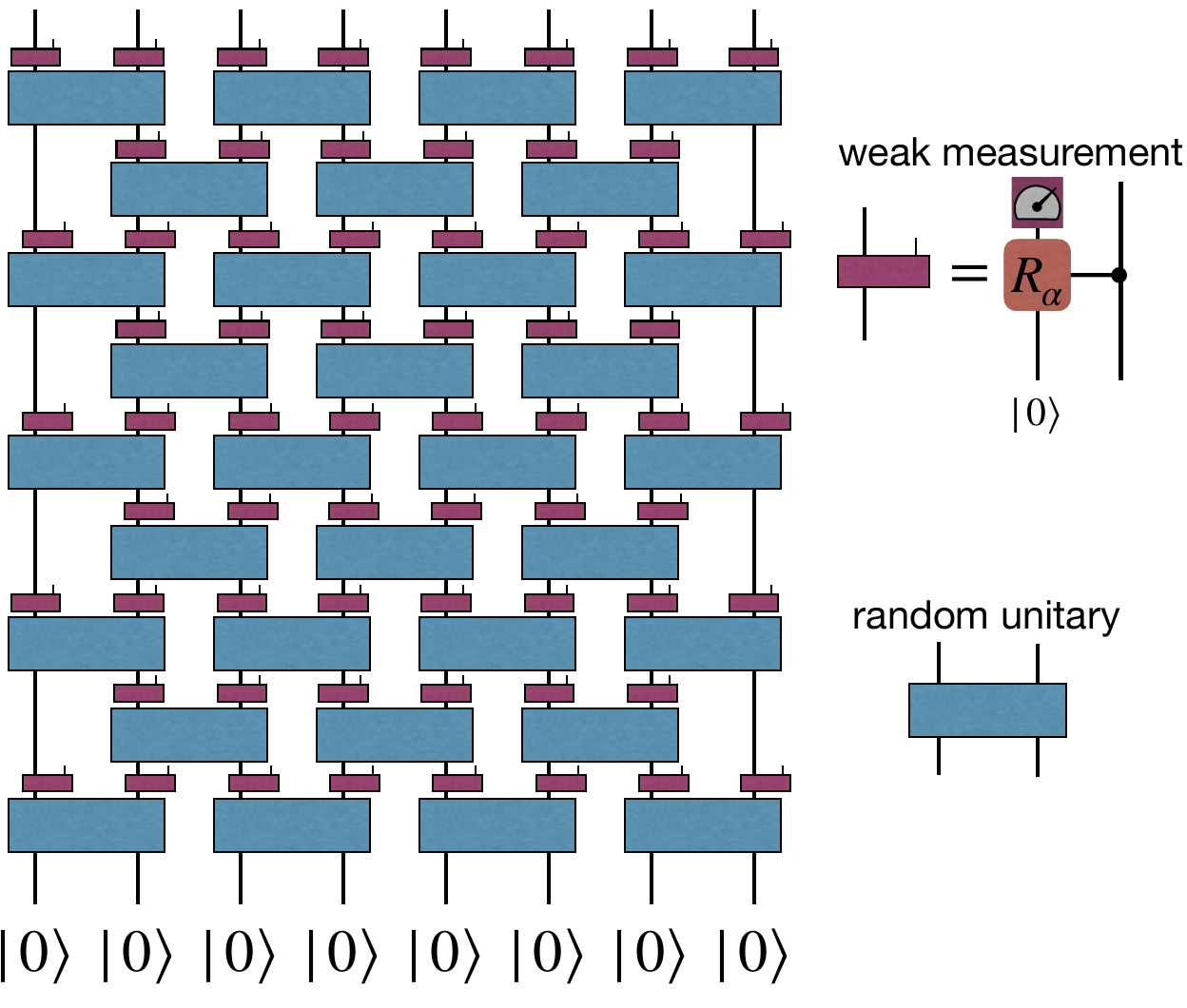}
    \caption{\textbf{Hybrid Quantum Circuit:} each qudit ($q$-levels system) in the brick wall circuit is subjected to unitary evolutions and weak measurements. The blue bricks represent two-qubit random unitary gates, while the other red bricks indicate weak measurements.}
\label{fig:WeakMeasurementEhud}
\end{figure}

At small values of $p$, broken bonds are limited to small, isolated clusters. Therefore, the minimal cut has to traverse through a number of unbroken bonds that scale proportionally to time. On the other hand, when $p$ exceeds a critical percolation threshold $p_c$, broken bonds predominate over unbroken bonds, forming a giant connected cluster of unbroken bonds that spans the entire system. As a result, the sole contribution to the entanglement arises from the bond breaking
near the starting point of the cut. Once this set of unbroken bonds is traversed, the minimal cut can extend arbitrarily far in the time direction using only broken bonds. Consequently, the average value of $S_{0,A}$ becomes independent of time at large $t$. Finally, if $p=p_c$ it can be shown that $S_{0,A}$ has a logarithmic scaling in time \cite{chayes1986critical}. In fact, one can argue that, if $R_j$ is the linear size of the $j$-th cluster of broken bonds in which the minimal cut passes, it can be shown that $R_{j+1}/R_j \sim 1 + \delta > 1$ on average and typically the path goes from $R_j$ to $R_{j+1}$ in $\mathscr{O}(1)$ time. As a consequence, $t \sim \sum_{j=1}^n R_j \propto (1 + \delta)^n$, while $S_{0,A} (t,p_c) \propto n$. Therefore, eliminating $n$ from the equation,
\begin{equation}
    S_{0,A} (t,p_c) \propto  \ln t \, .
\end{equation}
Note that the problem of identifying the most cost-effective path in a disordered medium, which is equivalent to the computation of $S_{0,A}$, has been extensively explored in mathematical literature and is often referred to as \textit{first passage percolation} \cite{chayes1986critical}. 


The emergence of a directed percolation critical point in the previous section was based on the use of $S_{0,A}$ to quantify entanglement. Indeed, in this case, the cost of a bond for the minimal cut cannot exhibit any fluctuations and the only relevant information is whether the bond was broken by a measurement or not. For more refined entanglement entropies $S_{n,A}$, the individual fluctuations of the cost of each bond play a role in determining the critical behaviour, which is thus different from the simple directed percolation. As we will see in the next section, directed percolation can once again be recovered in the limit of a large local Hilbert space dimension in the Haar group, so that once again fluctuations in the cost of each bond can be suppressed.

\subsection{Mapping to Spin Models}

\begin{figure}
    \centering
    \includegraphics[width=0.5\linewidth]{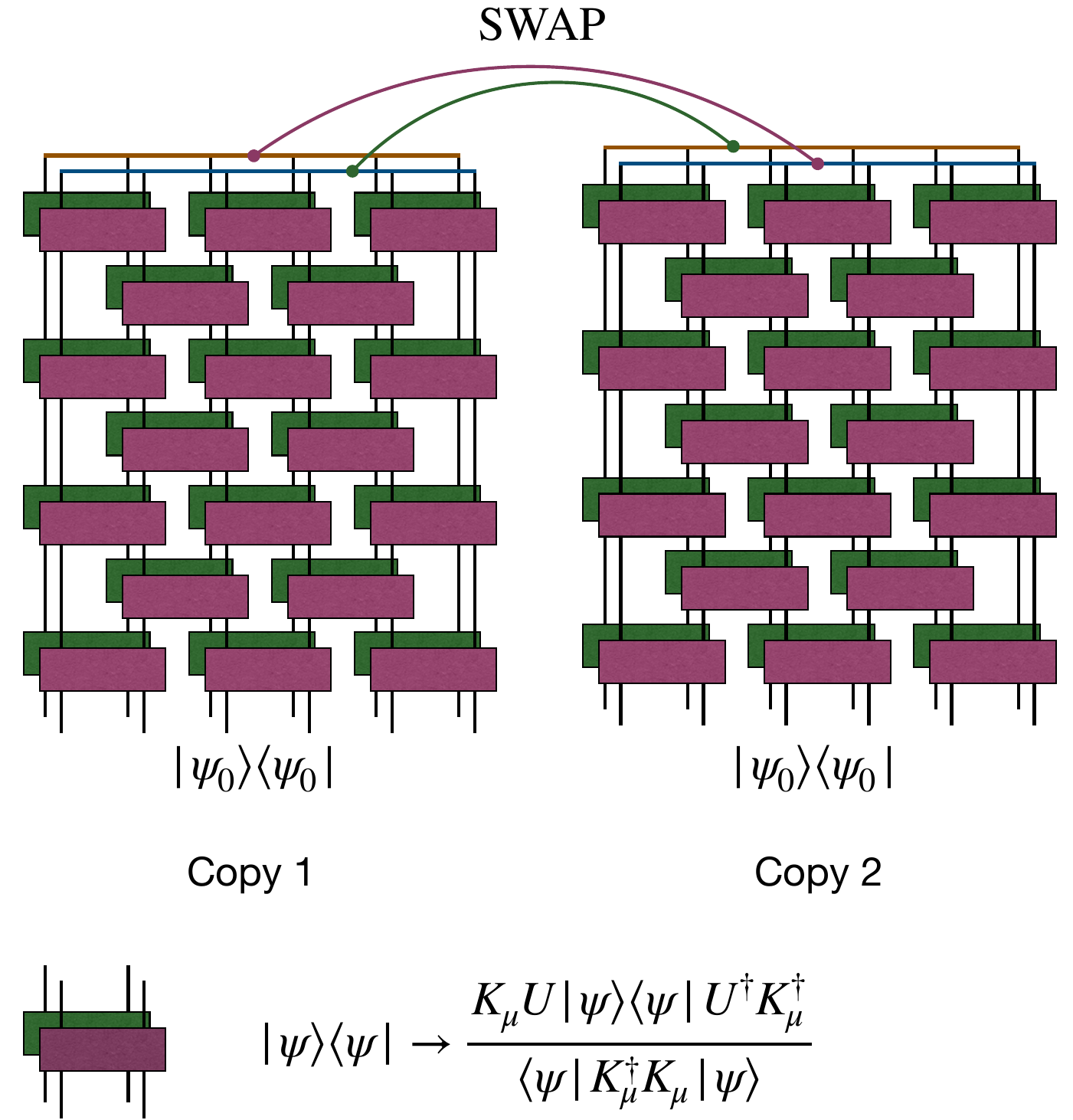}
    \caption{\textbf{SWAP Operator:} contractions in the replicated quantum circuit in order to compute a SWAP contraction. The blocks in the folded circuit represent the action of a unitary evolution followed by a weak measurement. The SWAP operator $C^{(2)}$ connects the indices of two replicas in order to map $\Tr{\rho^2}$ into $ \Tr{C^{(2)} \left(\rho\otimes \rho\right)}$.}\label{fig:swap_replicas}
\end{figure}
\begin{figure*}[t]
    \centering
    \includegraphics[width=0.7\linewidth]{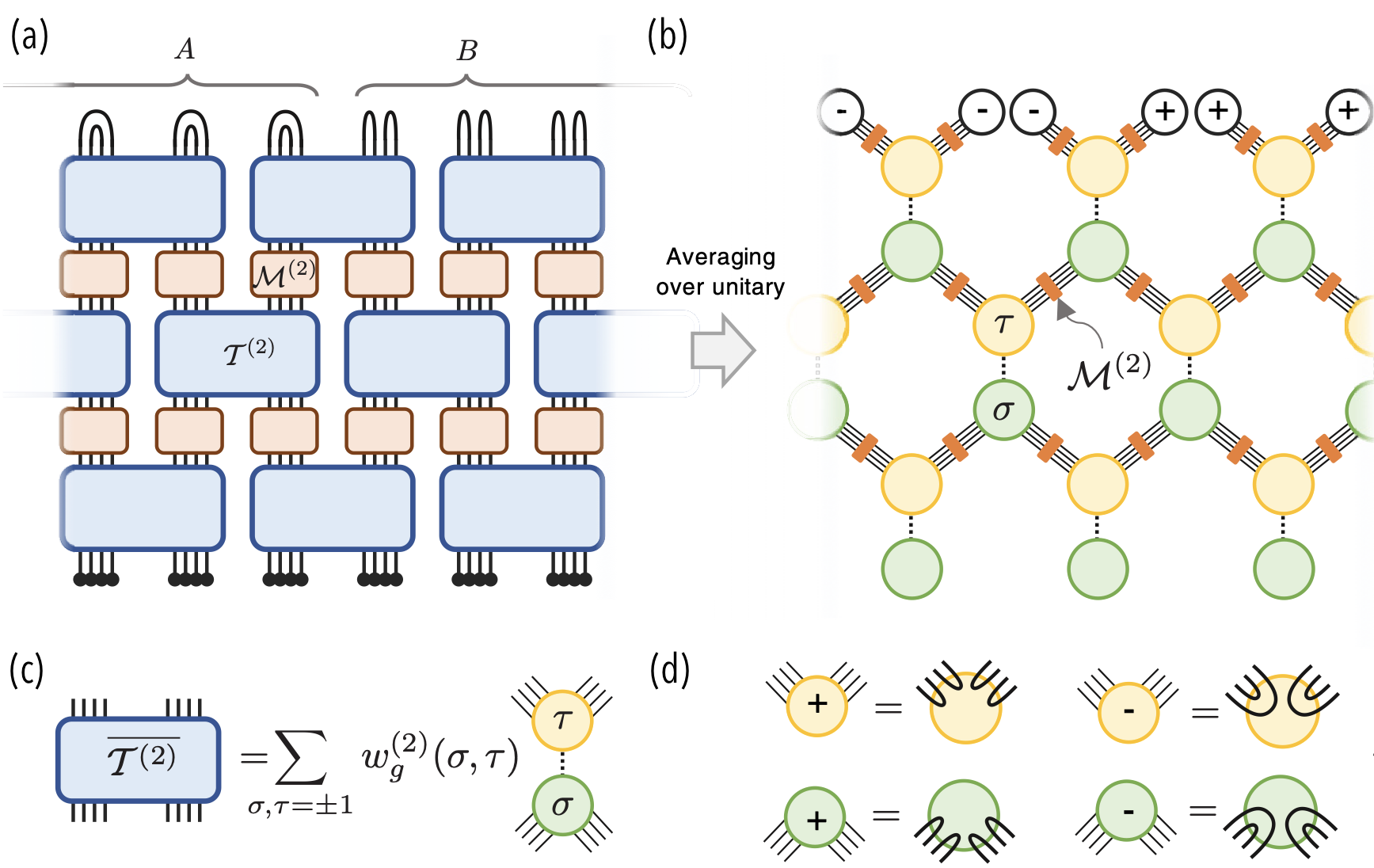}
    \caption{\textbf{Mapping to a spin model}. Figure adapted from \cite{Altman2020PRB}. (a) Tensor network representation of the RUC with weak measurements. In the fold-replicated circuit, each $T^{(2)}$ represents four identical random unitary tensors, and $M^{(2)}$ represents a weak measurement. (b) After averaging over unitaries, the tensor network is reduced to a network on a honeycomb lattice. Each site of the lattice hosts a classical Ising variable $\sigma$ or $\tau$. (c) Averaging $T^{(2)}$ over Haar random unitary allows one to replace the tensor as a sum of simple diagrams labelled by a pair of classical variables $\sigma, \tau \in {\pm 1}$. (d) Diagrammatic representations of the tensors $\hat{\tau}$ and $\hat{\sigma}$.
}
    \label{fig:2design}
\end{figure*}
{

We finally examine the work of Bao et al. \cite{Altman2020PRB} in which a similar mapping with the classical statistical model is developed. In particular, the random unitary circuit considered is made out $q$-levels systems (qudits), while a weak measurements protocol is considered, in which each qudit is coupled with an ancilla $M_j$, subject to projective measurements (see Fig. \ref{fig:WeakMeasurementEhud}). The random unitary local gate of the circuit in Fig.~\ref{fig:WeakMeasurementEhud} is being sampled from the Haar Measure on the unitary group $U(q^2)$. The Haar measure is actually the uniform measure over the group.

It is convenient to consider the ancillas as part of our system: in this case Eq. \eqref{eq:average_SRUC} can be reformulated as the conditional entropy of subsystem $A$ with its ancillas
\begin{equation}
    S_A = \overline{ \tilde{S}(A|M)} = \overline{{S} (\tilde\rho_{AM})}- \overline{{S} (\tilde\rho_{M})} \, 
\end{equation}
$\tilde{\rho}_X$ being the entanglement entropy of subsystem $X$ when the ancilla qudits are projected onto their computational basis $\ket{\mathbf{i}_M}$, i.e. $\tilde\rho = \sum_{\mathbf{i}} \dyad{\mathbf{i}_M} \expval{\rho}{\mathbf{i}_M}$. 
In turn, $S_{AM}$ can be computed through the replica formalism by introducing $n$ copies of the system
\begin{equation}
\label{ncopiesrho}
    \tilde{\rho}_{AM}^{(n)} = \tilde{\hat{\rho}}_{AM,1}\otimes ... \otimes \tilde{\hat{\rho}}_{AM,n} \, ,
\end{equation} 
One practically performs the calculation in the framework of $n-$copies, which simplifies the calculation of the trace from a logarithmic function to a polynomial one of the density matrix, and then we employ the analytic continuation to $n=1$
\begin{equation}
\label{eq:SA}
    S_A = \lim_{n \rightarrow 1} \tilde{S}^{(n)}(A|M) = \lim_{n \rightarrow 1}  \frac{1}{1-n} \left( \log \bar{\tr \rho^n_{AM} } - \log \bar{\tr \rho^n_{M} } \right) \, .
\end{equation}
As customary the $n$-th moments are represented by the $n-$replicated density matrix e.g. $\rho^n_{AM}$ and its trace  can be computed as 
\begin{equation}
   \tr \rho^n_{AM}  = \tr{\mathcal{C}^{(n)}\tilde{\rho}_{AM}^{(n)} }.
\end{equation}
where $\mathcal{C}^{(n)}$ represents an appropriate permutation operator and in particular it permutes every copy $i \to (i+1) \mod n$. Finally, as the evolution of $\rho$ contains both $K_\mu$ and $K_\mu^{\dagger}$ (weak measurement Kraus operators), which can be seen as a forward and a backward evolution, one has to consider $2 n$ folded copies (see Fig. \ref{fig:swap_replicas} for $n=2$). 
Averaging a random gate over the Haar measure \cite{Mele2024introductiontohaar} can be seen, heuristically, as a contraction of the indices of the $n$ copies of the random gate $U$ with the indices of the $n$ copies of $U^\dagger$, coming from the backward evolution. This is directly expressed by the average of the $2n$ copy tensor:
\begin{equation}
\label{eq;2ncopyaverage}
    \mathbb{E}_{\text{Haar}}[U^{n \otimes}\otimes U^{\dagger n \otimes}]= \sum_{\sigma,\tau \in \mathcal{S}_n} w_g^{(n)}(\sigma,\tau) \mid \tau \rangle \rangle \langle \langle \sigma \mid
\end{equation}
where $\mathcal{S}_n$ is the symmetric group of degree $n$ and, practically, contains all the possible permutations of $n$ objects.This tensor is a matrix acting on a $(q^2)^{2n}$ dimensional Hilbert space, since $U$ acts on 2 qudits. If we denote this Hilbert space  as $\mathcal{H}_n$, then the  vectors $\mid \sigma \rangle \rangle$  
 (with explicit components, $\braket{a_1, \ldots, a_n, \bar{a}_1, \ldots,  \bar{a}_n}{\sigma}\rangle = \prod_{i=1}^n \delta_{a_i, \bar{a}_{\sigma_i}}$) on $\mathcal{H}_n$, are the ones performing the permutation $\sigma$, by contraction of the degrees of freedom of the $U$ copies to the ones of the $U^\dagger$ copies. A graphical representation of this is demonstrated in Fig.~\ref{fig:2design}c.
More precisely, every pair of permutations appearing in Ea. \eqref{eq;2ncopyaverage}, must be weighted by the so-called  Weingarten function $ w_g^{(n)}(\sigma,\tau)$ \cite{köstenberger2021weingarten}. Without going into the technical details, the main result of\,\cite{Altman2020PRB} is that, after the averaging over the random 
unitary, Eq. \eqref{eq;2ncopyaverage} suggests that  each gate can be associated with two classical spin variables $\sigma$ and $\tau$,which can assume $n!$ values, and effectively interact through the measurements, giving rise to a classical model on a honeycomb lattice (see Fig.\,\ref{fig:2design}c)). In turn, this can be reduced to a spin model on a triangular lattice by summing over the 
$\tau$ variables. The mapping is actually possible only for large enough (but finite) $q$, since for small $q$ the emergent Boltzmann weights of the model can be negative. In particular, in the limit of $q \rightarrow \infty$ the spin model reduces to a $n!$-state nearest-neighbours Potts model. 

In this mapping, $\tilde{S}^{(n)}(A|M)$ corresponds to the difference in free energies relative to different boundary conditions: In particular, it can be seen as the cost of free energy associated with a domain wall that connects the two edges of $A$ at the boundary of the region. In the paramagnetic phase this continuation is $O(1)$, while it scales with the length $\ell_A$ of $A$ in the ferromagnetic region: these two different behaviours correspond respectively to the area law and the volume law of the model. 
To obtain the physical transition probability and the physical critical exponents, the $n \to 1$ limit has to be taken. As in this limit the Potts model reduces to a bond percolation model \cite{STEPHEN1976149}, we have $p_c = 1/2$ while $\nu = 4/3$ can be linked to the scaling of $S_A$ (see Eq.\,\eqref{eq:scalingSnu}).

In the Potts model, $n$ represents the number of spin states locally. Taking the $n \to 1$ limit is essentially letting the number of spin values to become trivially one. In this limit, the Potts model loses its ability to distinguish between different spin states, and only the occupancy of the lattice sites matters and that is why it reduces to a bond percolation model. In a bond percolation model, one only considers the existence or absence of connections (bonds) between lattice sites. The sites can be connected or not connected, with a probability p controlling the connection. There is also a critical probability ($p_c=1/2$) where a giant connected cluster first appears, signifying a percolation transition.

Beyond the large $q$ limit where the critical exponents of directed percolation are recovered, the full characterisation of the critical point of MIPTs at finite $q$ remains an open and intricate question. At the origin of this difficulty lies the necessity of the replica limit $n \to 1$ (see, e.g., \eqref{eq:SA}), which in turn requires analytical or at least explicit expressions.


\section{Partial local measurements affect critical correlations in quantum ground states}\label{sec:MIPT_in_GS}

In this section, we present some recent finding from \cite{garratt2022measurements}, where the authors  consider a simplified scenario to address the effect of quantum measurements on correlation functions. More specifically, they  consider a single shot of measurements in space on a given quantum state. Even in the absence of the interplay between measurements and unitary dynamics, they observe how measurements can drastically affect the long-distance behaviour of a strongly correlated quantum state. In particular, the focus is  on a critical ground state in 1D~\cite{PRXQuantumtremipt,miptentaglement}. This kind of state is strongly correlated, displaying algebraic decaying quantum correlations among local observables. Consequently, local measurements can wield substantial non-local effects. We will focus on Tomonaga-Luttinger liquid ground states, a continuous set of critical one-dimensional states labelled by the so-called Luttinger parameter $K$. The authors of  \cite{garratt2022measurements} prove that even exceedingly weak local measurements carried out across extensive spatial regions can alter the behaviour of long-distance correlations
while not entirely disentangling the degrees of freedom within the system. Consequently, the measured state remains notably correlated. To understand whether the measurement dynamics is relevant in the long-distance (infrared) regime, it is natural to look to the asymptotic properties of correlation functions.
\begin{figure}[h]
\centering
    \begin{subfigure}{0.4\textwidth}
    \centering
    \begin{tikzpicture}[x=0.75pt,y=0.75pt,yscale=-0.75,xscale=0.75]

\draw  [color={rgb, 255:red, 0; green, 0; blue, 0 }  ,draw opacity=0.2 ][fill={rgb, 255:red, 0; green, 119; blue, 255 }  ,fill opacity=0.35 ][line width=1.5]  (220,61) -- (420,61) -- (420,261) -- (220,261) -- cycle ;
\draw  [color={rgb, 255:red, 255; green, 255; blue, 255 }  ,draw opacity=0 ][fill={rgb, 255:red, 0; green, 119; blue, 255 }  ,fill opacity=0.6 ][line width=1.5]  (220,156) -- (420,156) -- (420,166) -- (220,166) -- cycle ;
\draw [color={rgb, 255:red, 0; green, 0; blue, 0 }  ,draw opacity=0.7 ][line width=1.5]  [dash pattern={on 5.63pt off 4.5pt}]  (220,159.5) -- (420,159.5)(220,162.5) -- (420,162.5) ;
\draw [line width=1.5]  (196,259.7) -- (440,259.7)(220.4,41) -- (220.4,284) (433,254.7) -- (440,259.7) -- (433,264.7) (215.4,48) -- (220.4,41) -- (225.4,48)  ;

\draw (196,149.4) node [anchor=north west][inner sep=0.75pt]  [font=\large]  {$\tau $};
\draw (307,263.4) node [anchor=north west][inner sep=0.75pt]  [font=\large]  {$x$};
\draw (285,63.9) node [anchor=north west][inner sep=0.75pt]  [font=\Large]  {$\phi ( x,\tau )$};

\end{tikzpicture}
    \caption{}
    \label{fig:singlereplica}
    \end{subfigure}
    \begin{subfigure}{0.4\textwidth}
    \centering
    \begin{tikzpicture}[x=0.75pt,y=0.75pt,yscale=-0.75,xscale=0.75]

\draw  [color={rgb, 255:red, 0; green, 0; blue, 0 }  ,draw opacity=0.2 ][fill={rgb, 255:red, 0; green, 119; blue, 255 }  ,fill opacity=0.35 ][line width=1.5]  (58,55.03) -- (258,55.03) -- (258,255.03) -- (58,255.03) -- cycle ;
\draw  [color={rgb, 255:red, 255; green, 255; blue, 255 }  ,draw opacity=0 ][fill={rgb, 255:red, 0; green, 119; blue, 255 }  ,fill opacity=0.6 ][line width=1.5]  (58,151.03) -- (258,151.03) -- (258,161.03) -- (58,161.03) -- cycle ;
\draw [color={rgb, 255:red, 0; green, 0; blue, 0 }  ,draw opacity=0.7 ][line width=1.5]  [dash pattern={on 5.63pt off 4.5pt}]  (58,154.53) -- (258,154.53)(58,157.53) -- (258,157.53) ;
\draw [line width=1.5]  (34,254.73) -- (278,254.73)(58.4,36.03) -- (58.4,279.03) (271,249.73) -- (278,254.73) -- (271,259.73) (53.4,43.03) -- (58.4,36.03) -- (63.4,43.03)  ;
\draw  [fill={rgb, 255:red, 0; green, 119; blue, 255 }  ,fill opacity=0.4 ][line width=1.5]  (276.12,68.2) .. controls (264.58,68.2) and (255.23,92.05) .. (255.23,121.47) .. controls (255.23,150.88) and (264.58,174.73) .. (276.12,174.73) .. controls (287.65,174.73) and (297,198.58) .. (297,227.99) .. controls (297,257.41) and (287.65,281.25) .. (276.12,281.25) -- (109.05,281.25) .. controls (120.58,281.25) and (129.93,257.41) .. (129.93,227.99) .. controls (129.93,198.58) and (120.58,174.73) .. (109.05,174.73) .. controls (97.52,174.73) and (88.17,150.88) .. (88.17,121.47) .. controls (88.17,92.05) and (97.52,68.2) .. (109.05,68.2) -- cycle ;
\draw [color={rgb, 255:red, 0; green, 0; blue, 0 }  ,draw opacity=0.7 ][line width=1.5]  [dash pattern={on 5.63pt off 4.5pt}]  (103.24,169.05) -- (267.82,167.77)(103.26,172.05) -- (267.84,170.77) ;
\draw [line width=1.5]    (100.36,73.3) .. controls (102.45,70.39) and (104.76,69.25) .. (108.28,67.43) ;
\draw [shift={(111.78,65.55)}, rotate = 150.56] [fill={rgb, 255:red, 0; green, 0; blue, 0 }  ][line width=0.08]  [draw opacity=0] (13.4,-6.43) -- (0,0) -- (13.4,6.44) -- (8.9,0) -- cycle    ;
\draw [line width=1.5]    (262.95,281.25) .. controls (265.34,281.28) and (270.84,281.29) .. (275.12,281.28) ;
\draw [shift={(279.08,281.25)}, rotate = 179.45] [fill={rgb, 255:red, 0; green, 0; blue, 0 }  ][line width=0.08]  [draw opacity=0] (15.72,-7.55) -- (0,0) -- (15.72,7.55) -- (10.44,0) -- cycle    ;
\draw  [color={rgb, 255:red, 0; green, 0; blue, 0 }  ,draw opacity=0 ][fill={rgb, 255:red, 0; green, 119; blue, 255 }  ,fill opacity=0.6 ][line width=1.5]  (97.46,165.25) .. controls (96.12,163.54) and (94.5,165.06) .. (101.65,165.16) .. controls (108.79,165.25) and (110.23,165.51) .. (131.02,165.65) .. controls (151.82,165.78) and (179.33,165.61) .. (203.18,165.65) .. controls (227.03,165.69) and (232.53,165.6) .. (243.26,165.65) .. controls (254,165.69) and (254.11,165.61) .. (259.15,165.65) .. controls (264.19,165.69) and (263.53,165.27) .. (263.92,165.65) .. controls (264.32,166.03) and (265.03,167.1) .. (265.78,168.52) .. controls (266.53,169.94) and (267.69,170.44) .. (268.61,171.65) .. controls (269.53,172.85) and (270.44,173.52) .. (272.11,174.27) .. controls (273.78,175.02) and (277.84,175.64) .. (279.72,175.65) .. controls (281.59,175.66) and (115.93,175.73) .. (114.25,175.65) .. controls (112.58,175.57) and (112.5,174.78) .. (110.79,174.68) .. controls (109.07,174.59) and (104.79,173.44) .. (103.46,172.49) .. controls (102.12,171.54) and (100.88,170.2) .. (99.65,168.49) .. controls (98.41,166.78) and (98.79,166.97) .. (97.46,165.25) -- cycle ;
\draw  [fill={rgb, 255:red, 0; green, 119; blue, 255 }  ,fill opacity=0.45 ][line width=1.5]  (317.18,82.92) .. controls (305.65,82.92) and (296.3,106.76) .. (296.3,136.18) .. controls (296.3,165.6) and (305.65,189.44) .. (317.18,189.44) .. controls (328.72,189.44) and (338.07,213.29) .. (338.07,242.71) .. controls (338.07,272.12) and (328.72,295.97) .. (317.18,295.97) -- (150.12,295.97) .. controls (161.65,295.97) and (171,272.12) .. (171,242.71) .. controls (171,213.29) and (161.65,189.44) .. (150.12,189.44) .. controls (138.58,189.44) and (129.23,165.6) .. (129.23,136.18) .. controls (129.23,106.76) and (138.58,82.92) .. (150.12,82.92) -- cycle ;
\draw [color={rgb, 255:red, 0; green, 0; blue, 0 }  ,draw opacity=0.7 ][fill={rgb, 255:red, 0; green, 119; blue, 255 }  ,fill opacity=0.45 ][line width=1.5]  [dash pattern={on 5.63pt off 4.5pt}]  (144.31,183.77) -- (308.89,182.48)(144.33,186.77) -- (308.91,185.48) ;
\draw [fill={rgb, 255:red, 0; green, 119; blue, 255 }  ,fill opacity=0.45 ][line width=1.5]    (141.43,88.01) .. controls (143.52,85.1) and (145.83,83.96) .. (149.35,82.14) ;
\draw [shift={(152.85,80.27)}, rotate = 150.56] [fill={rgb, 255:red, 0; green, 0; blue, 0 }  ][line width=0.08]  [draw opacity=0] (15.72,-7.55) -- (0,0) -- (15.72,7.55) -- (10.44,0) -- cycle    ;
\draw [fill={rgb, 255:red, 0; green, 119; blue, 255 }  ,fill opacity=0.45 ][line width=1.5]    (304.02,295.97) .. controls (306.41,296) and (311.9,296) .. (316.19,295.99) ;
\draw [shift={(320.15,295.97)}, rotate = 179.45] [fill={rgb, 255:red, 0; green, 0; blue, 0 }  ][line width=0.08]  [draw opacity=0] (15.72,-7.55) -- (0,0) -- (15.72,7.55) -- (10.44,0) -- cycle    ;
\draw  [color={rgb, 255:red, 0; green, 0; blue, 0 }  ,draw opacity=0 ][fill={rgb, 255:red, 0; green, 119; blue, 255 }  ,fill opacity=0.45 ][line width=1.5]  (138.52,179.97) .. controls (137.19,178.25) and (135.57,179.78) .. (142.71,179.87) .. controls (149.86,179.97) and (151.3,180.23) .. (172.09,180.36) .. controls (192.89,180.5) and (220.4,180.32) .. (244.24,180.36) .. controls (268.09,180.4) and (273.59,180.32) .. (284.33,180.36) .. controls (295.07,180.41) and (295.18,180.32) .. (300.22,180.36) .. controls (305.26,180.4) and (304.59,179.98) .. (304.99,180.36) .. controls (305.39,180.74) and (306.09,181.82) .. (306.84,183.23) .. controls (307.59,184.65) and (308.76,185.16) .. (309.68,186.36) .. controls (310.59,187.57) and (311.51,188.23) .. (313.18,188.98) .. controls (314.84,189.73) and (318.91,190.35) .. (320.78,190.36) .. controls (322.66,190.37) and (157,190.44) .. (155.32,190.36) .. controls (153.64,190.28) and (153.57,189.49) .. (151.86,189.39) .. controls (150.14,189.3) and (145.86,188.16) .. (144.52,187.2) .. controls (143.19,186.25) and (141.95,184.91) .. (140.71,183.2) .. controls (139.48,181.49) and (139.86,181.68) .. (138.52,179.97) -- cycle ;

\draw (60,58.43) node [anchor=north west][inner sep=0.75pt]  [font=\large]  {$\phi _{0}$};
\draw (102.8,71.6) node [anchor=north west][inner sep=0.75pt]  [font=\large]  {$\phi _{1}$};
\draw (144.12,86.32) node [anchor=north west][inner sep=0.75pt]  [font=\large]  {$\phi _{2}$};
\draw (36,143.43) node [anchor=north west][inner sep=0.75pt]  [font=\large]  {$\tau $};
\draw (145,258.43) node [anchor=north west][inner sep=0.75pt]  [font=\large]  {$x$};

\end{tikzpicture}
    \caption{}
    \label{fig:multiplereplicas}
    \end{subfigure}
    \caption{The two different instances of the field theory formalism (a) A single, specific set of measurement outcomes $m$, where the field theory is being described by a scalar field $\phi(x, \tau)$ in the presence of the perturbation originating from the operators $\hat{K}_m$ at $\tau = 0$. The imaginary-time path integral of $e^{-\beta \hat{H}}$, over the different configurations of $\phi(x,\tau)$ is represented by the light-blue regions. The correlations are obtained in the time frame $\tau=0$ and the measurements act in the ground state, which implies focussing on the limit $\beta \to \infty$. (b) The non-linear averages over the possible outcomes $m$ are being calculated with the help of the replica trick, according to which one has to deal with a field theory of $n$ replicated fields $\{\phi_i\}_{i=1}^n$. The average over the measurement outcomes $m$ has the effect of coupling of the different “sheets” of the replicas at $\tau=0$.}
\end{figure}
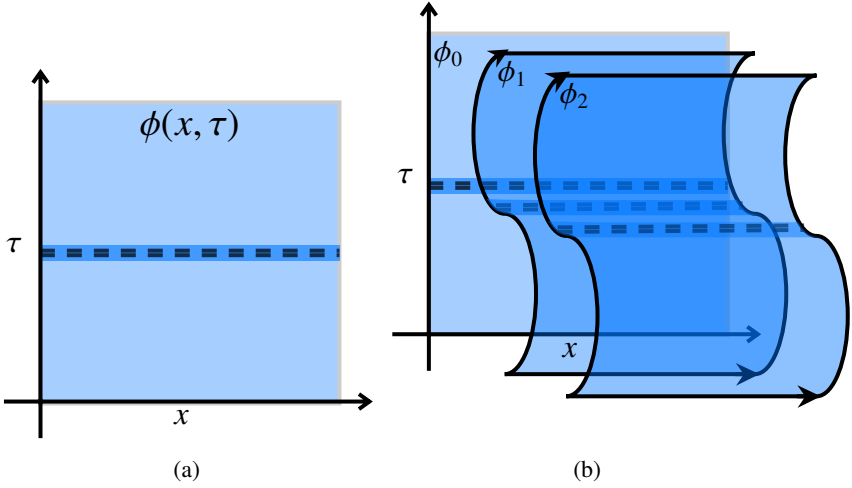
\subsection{Field Theory Description}
In this section, we will provide some generic field-theory tools that will be useful in the next section. Let us thus consider the ground state $\ket{\Psi_{GS}}$ of the Hamiltonian $\hat{H}$ and the relative projector $\ket{\Psi_{GS}} \bra{\Psi_{GS}}$. The latter can be seen as the limit $\beta \to \infty$, for thermal state,  $\rho \sim e^{-\beta \hat{H}}$ as in this regime only the leading contribution comes from the smallest energy mode is contributing.

In turn, the thermal state can be seen as the evolution operator in imaginary time $\tau$, between $\tau = 0$ and $\tau = \beta$. For example, the normalization of the state $e^{-\beta \hat{H}}$, i.e. the partition function, can be written in the path-integral formalism as
\begin{equation}
\label{eq:pathintegralfinitebeta}
\text{Tr} e^{-\beta \hat{H}} = \int \mathcal{D}\phi  \ e^{-S[\phi]}    
\end{equation}
where $\phi = \phi(x, \tau)$ represents a scalar field, eigenvalues of the quantum field $\hat{\phi}(x,\tau)$, and the action $S[\phi]$ involves an integral over spatial coordinates $x$ ($d$-dimensional) and the compact imaginary time interval $\tau \in [0, \beta]$. Since the partition function is given by a trace, this means that one has to impose periodic boundary conditions on the imaginary time direction, namely $\phi(x, 0) = \phi(x, \beta)$.

We will now consider the effect of a series of weak measurements on the ground state $\ket{\Psi_{GS}}$, described by the action of a set of Hermitian Kraus operators $\hat{K}_m $, $\sum_m K_m^2 = 1$ ($m$ corresponding to the measurement outcomes). One can leave for the moment the Kraus operators generic, and  specify them later in an example.
By taking into account the normalization one has thus that the state after the measurement protocol is given by $|\Psi_m\rangle = \hat{K}_m |\Psi_{GS}\rangle / \sqrt{p_m}$ where, as usual, the normalization $p_m$ corresponds to the Born probabilities 
\begin{equation}
\label{Bbornprobhamitlinan}
    p_m = \Tr(\dyad{\Psi_{GS}}{\Psi_{GS}} \hat{K}^2_m)= \lim_{\beta \to \infty} \frac{\Tr( \hat{K}_m^2 e^{-\beta \hat{H}})}{\Tr( e^{-\beta \hat{H}})}
\end{equation}


These weak measurements modify the amplitudes of the various contributions to the many-body state, but do not completely disentangle the system's degrees of freedom. As a result, the state post-measurement remains highly nontrivial and the important question that is being answered later is whether there is a change of the asymptotic properties of correlation functions with respect to the ground state. Following Ref.\,\cite{garratt2022measurements} we will focus on correlations over the states $\ket{\Psi_m}$, which in turn can be written as the expectation values $\expval{\cdot}_m$  over the density matrices
\begin{equation}
\label{densitymatpathint}
\begin{aligned}
    \dyad{\Psi_m} &= \frac{1}{p_m} \hat{K}_m\ket{\Psi_{GS}}\bra{\Psi_{GS}}\hat{K}_m \\ 
    &=\lim_{\beta \to \infty}\frac{\hat{K}_m e^{-\beta \hat{H}} \hat{K}_m}{\Tr \big(\hat{K}_m^2 e^{-\beta \hat{H}}\big)} \ . 
\end{aligned}
\end{equation}
Given a generic observable $\hat{O}$, we thus have that its expectation value can be rewritten in the path-integral formalism as 
\begin{equation}
\label{expvalpathint}
\langle \hat{\mathcal{O}} \rangle_m = \frac{\int \mathcal{D}\phi \ \langle \phi^\beta | \hat{K}_m \hat{\mathcal{O}} \hat{K}_m|\phi^0 \rangle \ e^{-S[\phi]}}{\int \mathcal{D}\phi \ \langle \phi^\beta | \hat{K}^2_m |\phi^0 \rangle \ e^{-S[\phi]}} \quad (2) \ . 
\end{equation}
For simplicity, let us introduce the notation $\phi^0 \equiv \phi(x, 0)$ and $\phi^\beta \equiv \phi(x, \beta)$. If one assumes that both $\hat{K}_m$ and $\hat{O}$ commute with $\hat{\phi}(x)$, it can be shown that the correct boundary conditions are still given by $\phi^\beta = \phi^0$. Moreover, in this case we have the freedom to translate in time $\hat{K}_m$, so that it can be chosen to act at $\tau = 0$. As $\hat{K}_m$ is assumed to be a local operator in space, the action of the measurements can then be seen as perturbations on the $d$-dimensional $\tau = 0$ surface in the $(d+1)$-dimensional field theory (see Fig.~\ref{fig:singlereplica}). The influence of the measurement on the long-distance structure of the ground state can now be addressed by determining whether these perturbations are relevant or not in  the Renormalization Group (RG) sense, with respect to the fixed point corresponding to the critical quantum ground state\,\cite{Hollowood2013}. 

 As anticipated, the evaluation of the average over the measurement outcomes $m$ of linear functionals of the state $\dyad{\Psi_m}$, weighted with the Born probabilities $p_m$,
 is equivalent to discarding the measurement outcomes. Thus, it generally leads to a loss of information, or complete dephasing towards infinite temperature, thus killing any non-local effects on the expected values of observables. This is demonstrated later in Eq. \eqref{linearaverfails} for the specific case of Tomonaga-Luttinger liquids. Interesting physics  emerges instead if one considers the average of non-linear quantities, e.g. $\braket{\Psi_m}{O | \Psi_m}^2$, over the distribution of measurement outcomes $p_m$, namely
 \begin{equation}
     \sum_m p_m\braket{\Psi_m}{O | \Psi_m}^2 \, .
 \end{equation}
To deal with this type of quantities, one uses that $\ket{\Psi_m} = p_m^{-1}\hat{K}_m \ket{\Psi_{GS}}$, so that
one can resort to the replica formalism using the replica trick
\begin{equation}
\label{eq:rpelicalimitfornonlinearaverage}
    \sum_m p_m\braket{\Psi_m}{\hat{O} | \Psi_m}^2=\lim _{n \rightarrow 1} \sum_m p_m^{n-2}\Tr(\hat {O}\hat{K}_m\dyad{\Psi_{GS}}{\Psi_{GS}} \hat{K}_m)^2
\end{equation}
for the non-linear averaging over measurement outcomes $m$. In the case of  continuous measurement outcomes, where $m$ becomes a scalar field $m(x)$, one can interpret the sum $\sum_m$ above as an integral. This technique allows for an exact calculation over multiple replicas $n$ and then by analytical continuation for a single replica as well, which is the case of interest. 
To develop the replica field theory, 
we recall that  $p_m = \Tr(\hat{K}_m\dyad{\Psi_{GS}}{\Psi_{GS}} \hat{K}_m)$, so that Eq.~\eqref{eq:rpelicalimitfornonlinearaverage} involves $n$ replicas, two of which involving the insertion of the operator $\hat {O}$. Then,
one uses Eqs. \eqref{Bbornprobhamitlinan},\eqref{expvalpathint}  to express all factors using the path-integral representation of $e^{-\beta \hat{H}}$ . Finally, one can exploit the path-integral formalism of  to express the quantity  \eqref{eq:rpelicalimitfornonlinearaverage} directly as a field theory of $n$ replica fields $\phi_i(x,\tau)$ ($i=0,1,\dots,n-1$). Within this framework, after the non-linear averaging, the replicas do not interact in the bulk, while they are coupled at the $\tau=0$ boundary by the sum over $m$, as shown in Fig.~\ref{fig:multiplereplicas}
In the following, according to  \cite{garratt2022measurements}, one can work in the zero-click limit (to be clarified afterwards), where the replica limit is not needed and thus one can focus for simplicity on a single replica field theory, implying that the effects of the measurements on the critical correlations do not depend qualitatively on the number of replicas.

\subsection{Application to Tomonaga-Luttinger liquids}
The framework described in the previous section has been applied in Ref.\,\cite{garratt2022measurements} to the Tomonaga-Luttinger liquids (TLLs)\,\cite{giamarchtll,FDMHaldane_1981} of one-dimensional spinless fermions. 
The Tomonaga-Luttinger liquid (TLL) Hamiltonian before bosonization is formulated in terms of fermionic operators. It captures the kinetic energy of right-moving and left-moving fermions near the Fermi points and their interactions. The Hamiltonian of this type of system  is given by:
\begin{equation}
H = \int dx^d \left[ \hat{\psi}_R^\dagger(x) (-iv_F \partial_x) \hat{\psi}_R(x) + \hat{\psi}_L^\dagger(x) (iv_F \partial_x) \hat{\psi}_L(x) \right] + \frac{1}{2} \int dx \, dx' \, \hat{\varrho}(x) V(x - x') \hat{\varrho}(x'),
\end{equation}

where:
\begin{itemize}
    \item  $\hat{\psi}_R(x)$ and $\hat{\psi}_L(x)$ are the right-moving and left-moving fermionic field operators,
\item $v_F$ is the Fermi velocity,
\item $\hat{\varrho}(x) = \hat{\varrho}_R(x)+\hat{\varrho}_L(x)=\hat{\psi}_R^\dagger(x) \hat{\psi}_R(x) + \hat{\psi}_L^\dagger(x) \hat{\psi}_L(x)$ is the total density operator,
\item $V(x - x')$ is the interaction potential between fermions.
\end{itemize}

This model can be mapped onto a free bosonic theory by means of the bosonization formalism\,\cite{giamarchtll}: in particular, the normal ordered particle density operator $\hat{n}(x)$ can be expressed in terms of the so-called \emph{counting field} $\hat{\phi}(x)$ as
\begin{equation}
\label{eq:densityfield}
    \hat{n}(x)=:\hat{\varrho}(x):=- \frac{1}{\pi} \nabla \hat{\phi}(x)+ \frac{1}{\pi} \cos \left( 2 k_F x-2 \hat{\phi}(x) \right) 
\end{equation}
where $k_F$ is the Fermi momentum and higher Fourier components (oscillating with higher frequencies $4k_F,6k_F,\dots$) have been neglected. The counting field plays the role of the quantum field $\hat{\phi}(x)$ introduced earlier on more general grounds.
\par In this case, the field-theoretical action (without the measurements) arising from the bosonization procedure takes the form
\begin{equation}
\label{eq:TTLaction}
S[\phi] = \frac{1}{2\pi K} \int d^d x \int_{0}^{\beta} d\tau \left[ \left(\frac{\partial \phi}{\partial \tau}\right)^2 + (\nabla \phi)^2 \right]  \,  . 
\end{equation}
The coupling constant $K$ (not to be confused with the Kraus operators introduced above) is called the Luttinger parameter and characterises the interaction of the fermions. Specifically, for $K=1$ the theory represents free fermions, whereas for $K>1, K<1$ it describes the fermions with attractive or repulsive interactions respectively.

Alternatively, the action $S$ can be written in terms of the field $\theta(x)$ associated with \emph{phase operator} $\hat{\theta}(x)$. Although $ \hat{\phi}$ is related to the density of fermions, $\pi^{-1} \ \nabla \hat{\theta}(x)$ is canonically conjugated to $\hat{\phi}$, $[\hat{\phi}(x), \nabla \hat{\theta}(x')] = i\pi \delta(x - x')$, and it is associated with phase fluctuations. In particular, the bosonization of LLT leads to the following relation
\begin{equation}
\begin{aligned}
\nabla \hat{\phi}(x) & =-\pi\left[\hat{\varrho}_R(x)+\hat{\varrho}_L(x)\right] \\
\nabla \hat{\theta}(x) & =\pi\left[\hat{\varrho}_R(x)-\hat{\varrho}_L(x)\right]
\end{aligned}
\end{equation}
with $\nabla \theta(x)$ counting the difference between right and left moving fermions and thus being the charge current operator. Equivalently, one can rewrite the action \eqref{eq:TTLaction} in terms of phase fluctuations: one finds that it follows a form similar to $S[\phi] $ with $\phi \rightarrow \theta$ and $ K \rightarrow K^{-1}$. 

As already pointed out, the ground state properties of the model emerge in the zero-temperature, $\beta \rightarrow \infty$, limit. In this regime, the free theoretical description \eqref{eq:TTLaction} foresees an algebraic decay in the correlations of the density $\hat{n}(x)$ and the phase $\hat{\theta}(x)$ 
\begin{equation}
\label{correlsnomeasure}
\begin{aligned}
 & \expval{\nabla \hat{\phi}(0) \nabla \hat{\phi}(x)}_{\text{GS}}  \sim x^{-2} \\
 & \expval{e^{i(\hat{\theta}(x) - \hat{\theta}(0))}}_{\text{GS}} \sim x^{-\frac{1}{2K}} \ . 
\end{aligned}
\end{equation}
 
\begin{figure}[t]
    \centering
    \input{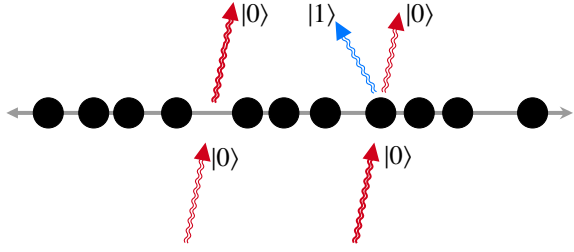}
    \caption{The ancilla qubit protocol. The ancilla qubits are initially prepared in the state $\ket{0}$ and afterwards, they weakly interact with the local particle density. The particles and the ancilla qubits are indicated with black disks and wavy arrows, respectively. In the case in which there is no particle ($n_j = 0$) the initial state of the ancilla qubit remains unchanged and the result of the measurement is $\ket{0}$, while when $n_j=1$ the ancilla qubit can be measured in either of the $\ket{0},\ket{1}$ states.}
    \label{fig:acillascheme}
\end{figure}
In the following sections, we present  the results from \cite{garratt2022measurements} on   how the local measurements can change the algebraic exponents of the critical correlations.

\subsection{Weak measurements with ancillary qubits}
We will now address how the presence of measurements changes the behaviour of the critical correlations between the ground state and the state where measurements have taken place. In particular, a way to perform weak measurements of the local densities is to use ancillary qubits coupled with $\hat{n}$. Let us analyse this protocol for a single site $j$: initially, one assumes the ancillary qubit to be in the eigenstate $\ket{0}$ of the Pauli operator $\hat{\sigma }_z$. Subsequently, it is coupled to the density operator, $\hat{n}_j$ through the Hamiltonian $H = u_j \hat{n}_j \otimes \hat{\sigma}^x$. The corresponding unitary operator after a time $\Delta t$ is given by
\begin{equation}
\label{couplingjanciall}
\begin{aligned}
\hat{\mathcal{U}}_j & =\exp \left[i \Delta t  u_j \hat{n}_j \otimes \hat{\sigma}^x\right] \\
& =1+ \left(\cos C_j -1\right) \hat{n}_j+i\sin C_j \hat{n}_j \otimes \hat{\sigma}^x \, , 
\end{aligned}
\end{equation}
where $C_j \equiv  u_j \Delta t$ and the second inequality is obtained by taking into account that $\hat{\sigma}_x^2=\mathrm{1}$ and that $\hat{n}_j^2=\hat{n}_j$ due to the fermionic nature of the model. After a time $\Delta t$, a projective measurement of $\hat{\sigma}_z$ on the ancillary qubit is performed. Since $\sigma_x \ket{0} = \ket{1}$, one has that the Kraus operators $\hat{K}_{0,1}$ corresponding to the outcomes $0$ and $1$ respectively are given by 
\begin{equation}
\label{measuredclicksnoclicks}
\begin{aligned}
& \hat{K}_0 = [\bra{0} \otimes \mathds{1}] \hat{\mathcal{U}}_j [\ket{0} \otimes \mathds{1}] = 1-\left(1 - \cos C_j \right) \hat{n}_j \,, \\
& \hat{K}_1 =[\bra{1} \otimes \mathds{1}] \hat{\mathcal{U}}_j [\ket{0} \otimes \mathds{1}] =\sin C_j \hat{n}_j \, ,
\end{aligned}
\end{equation}
so that when $n_j = 0$ only the outcome $m=0$ is possible, while when $n_j=1$ both outcomes are possible with a probability $\cos^2 C_j$ and $\sin^2 C_j$ respectively. Notice that $\hat{K}_m$ are hermitian and $[\hat{K}_m,\hat{n}_k] = 0$ for every site $k$. 
As a consequence, one has that the particle density correlation $\expval{\hat{n}_{0}\hat{n }_{x}}$ is preserved by the measurement protocol since, after the projective measurement of the ancilla, indeed
\begin{equation}
\label{linearaverfails}
\begin{split}
    \expval{\hat{n}_{0}\hat{n }_{x}}_{t = \Delta t} &= \sum_m p_m \frac{\ev{\hat{K}_m \hat{n}_{0}\hat{n }_{x} \hat{K}_m}{\Psi_{\text{GS}}}}{\ev{\hat{K^2}_m} {\Psi_{\text{GS}}}} \\ &= \sum_m \ev{\hat{n}_{0}\hat{n }_{x} \hat{K^2}_m} {\Psi_{\text{GS}}} = \expval{\hat{n}_{0}\hat{n }_{x}}_{GS} \ . 
\end{split}
\end{equation}
As anticipated, linear averages of correlation functions of the operator undergoing measurements are completely insensitive to the effects of measurement.
\par We will now generalise this protocol to the case in which an extensive amount of measurements is performed over space. To make things more straightforward, we focus on the no-click limit, in which only the trajectory with the outcome $m=0$ is retained. This procedure is effectively based on postselecting only those measurement outcomes where $m = 0$ and will in general be exponentially demanding in the number of repetitions.
The resulting state is, thus,
\begin{equation}
\label{noclicksmeasu}
\begin{aligned}
\left|\Psi_{\mathrm{0s}}\right\rangle \propto \hat{K}_{\mathrm{0s}}\left|\Psi_{\mathrm{GS}}\right\rangle \, , \\
\hat{K}_{\mathrm{0s}} & \equiv \prod_j\left[1- \left(1-\cos C_j\right) \hat{n}_j\right] \, . 
\end{aligned}
\end{equation}
with $\ket{\Psi_\mathrm{0s}}$ is the post-selected state of no-clicks. Moreover, one can consider the limit in which $C_j \ll 1$, in which  $\cos(C_j) = 1 -C_j^2/2 + O(C_j^4)$ and consequently,
\begin{equation}
    1- \left(1-\cos C_j\right) \hat{n}_j = e^{-  C_j^2 \hat{n}_j/2 + O(C_j^4)} .
\end{equation}
where one can use $\hat{n}_j^2=\hat{n}_j$ to prove that $e^{\lambda \hat{n}_j}=1+ [1-\cos(2\lambda)-1]\hat{n}_j$. As a consequence, $\hat{K}_{\text{0s}}$ becomes
\begin{equation}
\label{noclickcontinuouslimit}
    \hat{K}_{\text{0s}} = \exp{-\frac{1}{2} \sum_j C_j^2 \hat{n}_j} \propto \exp{-\frac{1}{2} \int dx \, C^2(x) \hat{n}(x)} . 
\end{equation}
Let us notice that $[\hat{K}_{\text{0s}},\hat{n}(x)]=0$, so that the density-density correlations over $\ket{\Psi_{\mathrm{0s}}}$ can be expressed as 
\begin{equation}
\label{correlatornn}
\langle\hat{n}(0) \hat{n}(x)\rangle_{\mathrm{0s}}=\frac{\operatorname{Tr}\left[e^{-\beta \hat{H}} \hat{K}_{\mathrm{0s}}^2 \hat{n}(0) \hat{n}(x)\right]}{\operatorname{Tr}\left[e^{-\beta H} \hat{K}_{\mathrm{0s}}^2\right]} \, .
\end{equation}
In turn, once Eq.\,\eqref{correlatornn} is expressed in the path-integral formalism and, since one is interested in correlations at $\tau=0$, one can integrate out the fluctuations of the scalar field $\phi(x, \tau)$ in Eq.\,\eqref{eq:TTLaction} for $\tau \neq 0$. By using the notation, $\phi(x,0)\to \phi(x)$ from now on and the Fourier  transform of the $\tau=0$ field $\tilde{\phi}(k)=\int dx e^{-ikx} \phi(x) $, one can obtain the numerator of Eq. \eqref{correlatornn} as
\begin{equation}
\label{eq:measureactionperturn}
\int \mathcal{D}\varphi \ e^{-S_{\tau=0}[\phi] -  \int dx \ C(x)^2 \ n(x) }n(0) n(x)
\end{equation}
where 
\begin{equation}
\label{eq:actiontau0}
    S_{\tau=0}[\phi]=\frac{1}{\pi K}\int \frac{dk}{2\pi} |k| \  |\tilde{\phi}(k)|^2
\end{equation}
is a non-local action encapsulating the $\tau=0$ contributions.  Mathematically speaking, the Fourier transform of $\tilde{\phi}(k)$ in Eq. \eqref{eq:actiontau0} leads to contributions of the form, $\phi(x)\phi(x')$ which is an indicator of non-locality. In principle, this action leads to the same correlations as $S[\phi]$. Additionally, the measurements, through  Eq. \eqref{noclickcontinuouslimit} result in the second term of the exponent in Eq. \eqref{eq:measureactionperturn}, which acts as perturbation to $S_{\tau=0}[\phi]$ and it depends on the coupling $C(x)$, with the weak measurement setup. This perturbed action is essentially representing an effective $1$-dimensional theory of our system under measurement and is defined as follows
\begin{equation}
\label{effectiveact}
    S_{\text{0s}}[\phi ] \equiv  S_{\tau=0}[\phi] - \int dx \ C(x)^2 \ n(x)
\end{equation}

Finally, one needs to express $n(x)$ in terms of the scalar field $\phi(x)$: by resorting to the operator identity \eqref{eq:densityfield}. However, before doing that, we should mention the interesting results occur from the second term in Eq. \eqref{eq:densityfield}, which oscillates with the wave number $2k_F$. In particular, this is demonstrated by knowing that the Fermi momentum fixes the mean interparticle separation to $\pi/k_F$. Thus, one can assume  that  weak measurements are being performed on locations $x=\alpha \pi/k_F, \quad \alpha \in \mathbb{Z}$, for which  one gets  post-measured states  described  by the action 
\begin{equation}
\label{effectiveact}
    S_{\text{0s}}[\phi ] \equiv  S_{\tau=0}[\phi] - C \int dx \cos(2\phi)
\end{equation}
where  $C$ is a constant that comes from the coupling strengths $C^2(x)$. 

Before going on, let us provide a more intuitive insight into the integration of $\tau \neq 0$ fluctuations. Indeed, $S_{\tau=0}[\phi]$ can be thought as the action of a local degree of freedom coupled to a zero-temperature Ohmic bath \cite{PhysRevLett.46.211}. When 
integrating out the physical bath for $x \neq 0$, an action emerges for a local degree of freedom at $x = 0$ that exhibits non-locality in time, with the non-locality encapsulating the memory of the bath. Analogously, in the case studied in this section, the roles of time and space are reversed, with the 'bath' now being represented by fluctuations of the field $\phi(x, \tau)$ at $\tau \neq 0$ and the final action $S_{\tau=0}[\phi]$  is now non-local in space. The nonlocality here conceals the spatial memory, which is the entanglement present in the ground state.

\subsection{Renormalization Group scheme}
Eq.\,\eqref{effectiveact}, which represents the effective action of the problem in the no-click limit of the post-measured case, corresponds to the celebrated Sine-Gordon field theory,\cite{giamarchtll}. It is known that the large distance behaviour of the theory can be obtained though a perturbative RG calculation in $C$. In particular, by integrating the fast fluctuation and sending $x \to e^{-l}x$ one finds that the coupling constant $C$ is renormalised according to the flow equation at small $C$ \cite{giamarchtll}
\begin{equation}
\label{flowofC}
    \dv{C}{l}= (1-K)C
\end{equation}
When $K>1$, the effects of the measurement are irrelevant, since the effective coupling $C$ decreases to zero. As a consequence, $\expval{\nabla \hat{\phi}(0) \nabla \hat{\phi}(x)}_{\text{0s}}$ and phase correlations $\expval{e^{i(\hat{\theta}(x) - \hat{\theta}(0))}}_{\text{0s}}$ exhibit the same large distance behaviour of the unmeasured ground state, given by Eq.\,\eqref{correlsnomeasure}. 

When, $K<1$ instead, $C$ flows to $+\infty$, which means that the effect of the measurement is relevant and the system flows toward the $C \gg 1$ regime. In this regime,  the kinetic term in the sine-Gordon will be negligible with respect to the potential term, and one obtains a saddle point approximation according to which  the path integral is dominated by the configuration of the field $\phi(x)$ which minimizes $\cos 2\phi$, i.e. a discontinuous function which jumps between the values $\phi = j\pi$, with $j \in \mathbb{Z}$. An example is presented in Fig.~\ref{fig:scalarstep}. Since the focus is solely on correlations over length scales much greater than the typical separation of the points of discontinuities, one can write
\begin{equation}
\label{philargeC}
\begin{aligned}   
&\phi(x) \approx \pi \sum_j \alpha_j \Theta(x-x_j) \\
& \nabla \phi(x) \approx  \pi \sum_j \alpha_j \delta(x-x_j)
\end{aligned}
\end{equation}
where $\Theta(x)$ is the step function, $\alpha_j=\pm 1$ and $x_j$ are the locations of the discontinuities. In fact, every configuration of $x_j$ corresponds to a different saddle point, but the dilute configurations are the most
dominant. By substituting the approximation of $\phi(x)$ from Eq. \eqref{philargeC} into $S_{\tau=0}[\phi]$, one obtains  a  logarithmic interaction between $x_j$ of the form  $2K^{-1} \sum_{i<j} \alpha_i \alpha_j \log|x_i-x_j|$ indicating a repulsion between discontinuities with the same $\alpha_j$ and attraction with the opposite $\alpha_j$. The coarse graining of the RG scheme modifies the length scale by a factor of \( e^l \), and eliminates oppositely signed $x_i,x_j$ with separations $|x_i - x_j| \leq e^l$. When \( K \) is sufficiently small, resulting in a strong attraction between such points, this process leads to the domain walls becoming increasingly dilute and in the regime of $C \gg 1$ the RG flow is described by
\begin{equation}
      \dv{C}{l} \propto \left( \frac{1}{K}-1 \right) C^{1/2}
\end{equation}
In the strong measurement limit, the field theory under $S_{\text{0s}}[\phi]$ and  considering \eqref{philargeC}, leads to the following correlations 
\begin{equation}
\label{correlaftermeasure}
\begin{aligned}
 & \expval{\nabla \hat{\phi}(0) \nabla \hat{\phi}(x)}_{\text{0s}}  \sim x^{-2/K} \\
 & \expval{e^{i(\hat{\theta}(x) - \hat{\theta}(0))}}_{\text{0s}} \sim x^{-\frac{1}{K}}
\end{aligned}
\end{equation}
In summarizing, we have that when \(K < 1\), the behaviour at long scales is characterized by dilutely  distributed points of a step scalar $\phi$ with correlations with different algebraic exponents. Comparing the  unmeasured correlations in Eq. \eqref{correlsnomeasure} with the ones in the measured case in Eq. \eqref{correlaftermeasure},  we can observe that correlations of $\nabla \hat{\phi}(x)$ decay faster, whereas phase fluctuations decay slower. This is a proof that arbitrarily weak measurements  suppress density fluctuations and  enhance phase fluctuations.
Consequently, for critical quantum ground states, which exhibit algebraic correlations between observables, the impact of local measurements is particularly significant at long distances, indicating their non-local effects and significant changes in the entanglement structure of the critical state.

On the other hand, for \(K > 1\), the measurements have no effect on large  distances and  fail to change the behaviour of correlations. Hence, there exists a transition in the response of the quantum state to measurement at \(K = 1\).

\begin{figure}[b]
    \centering
    \includegraphics[width=0.6\textwidth,keepaspectratio]{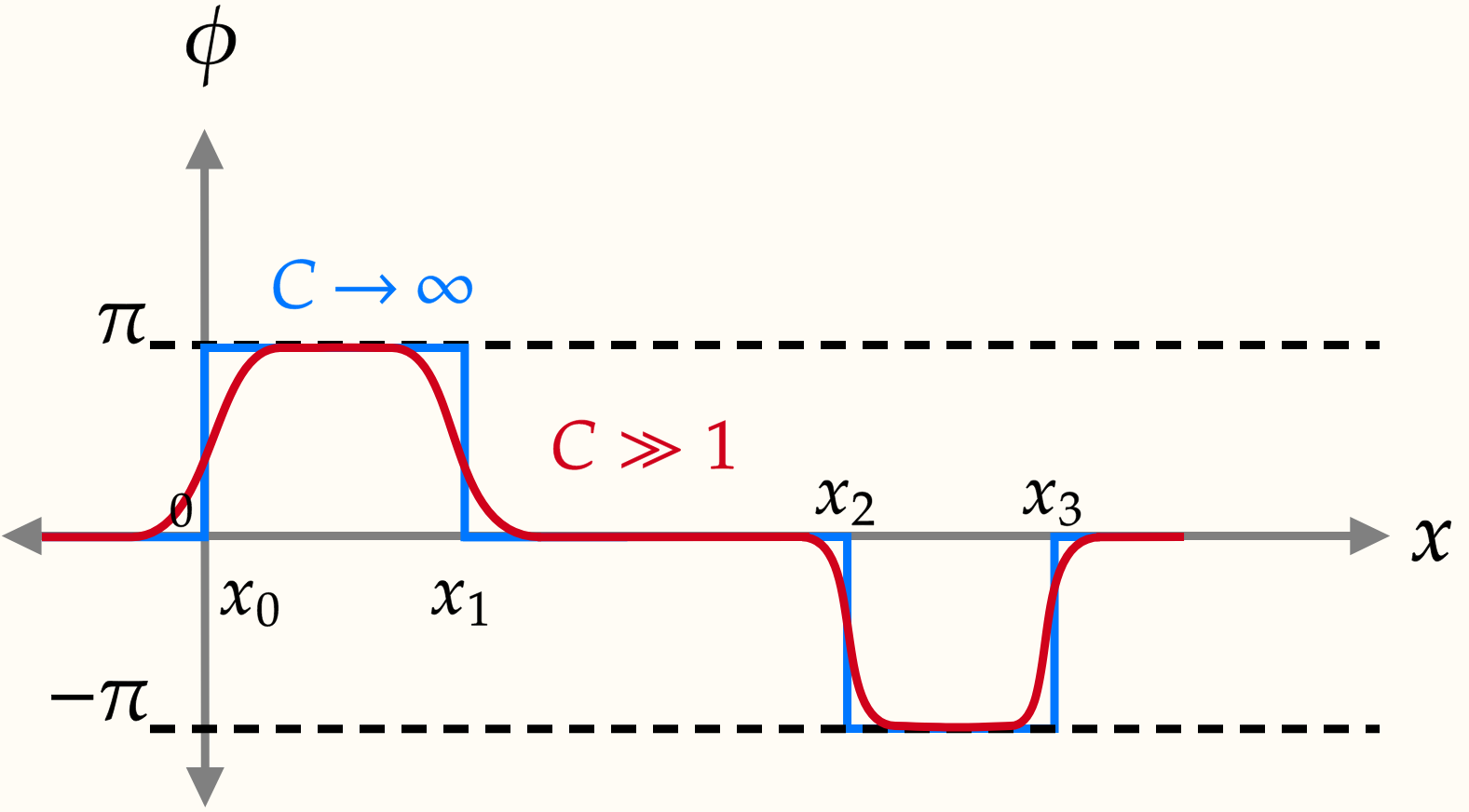}
    \caption{As the coupling strength $C$ increases, the saddle point approximation suggests that the field theory will be dominated more and more by scalars $\phi(x)$ who look like a step function, with discontinuities at some configuration of points $x_j$.}
    \label{fig:scalarstep}
\end{figure}

The above study was done for weak measurements, which were defined to be Hermitian and non-unitary local operations. In the case where the weak coupling is local unitary, the critical ground state is locally perturbed by unitary operations. The research done by the authors in \cite{cianciaruso2015entanglement} highlights how local unitary transformations, unlike global unitary perturbations, typically preserve local properties and do not inherently induce non-local changes unless specifically designed. This supports the idea that unitary perturbations maintain coherence and entanglement structure, contrasting with the non-local effects of measurements. 

\section{Conclusions and Open questions about MIPT}

In conclusion, this report has explored some features of measurements in quantum mechanics and their influence on the behaviour and properties of quantum systems. The focus was on understanding the relationship between measurements and entanglement, in the context of MIPT in the case of noisy quantum dynamics. Another focal point of the report was the coordinated influence of local measurements conducted on critical quantum ground states, leading to transitions in correlations at long distances. This drastic change in the behaviour of the quantum state arises from the intrinsic nonlocality embedded within the measurement process, as well as from the algebraic decay of correlations, characteristic of critical states.
A key outcome of our investigation is the clarification of how the non-local repercussions of measurements find explication through conventional tools within the realm of quantum statistical mechanics, such as path integral formalism and RG theory. Beyond \cite{garratt2022measurements}, these tools proved to be fruitful for analytical results in Brownian Sachdev-Ye-Kitaev (SYK) chains \cite{PhysRevLettSYKMIPT}, where the entanglement transition caused by continuous monitoring, is due to symmetry breaking in the enlarged replica space.
It is important to mention that in systems studied in \cite{giachetti2023elusive,PhysRevLettminato} there is an absence of MIPT. In particular, in the former paper, the authors present an exact solution for a system of $N$ spin $1/2$ particles with pairwise all-to-all noisy interactions, where the continuous isotropic monitoring reveals no many-body-induced phase transition. In the later paper, the authors consider the robustness of the MIPT for long-range interaction that decays as $x^\alpha$ with distance $x$ and showcase the absence of the transition, when the critical correlations decay slow enough $\alpha< \alpha_c$, even for arbitrarily strong measurements.

However, it is noteworthy that experiments on measurement-induced criticality often necessitate post-selection to unveil the underlying physics concealed in quantum trajectories. Although post-selection is a powerful tool for isolating and studying specific quantum effects, it introduces several experimental challenges that can affect the reliability and generalisability of the results \cite{Aaronson2005,Preskill2018quantumcomputingin}. Firstly, post-selection significantly reduces the effective size of the dataset, since only a fraction of all measurement outcomes are retained for further analysis. Secondly, the reliance on post-selection can also increase the complexity of experiments and the demand for high-fidelity quantum control and measurement, since it is practically challenging to condition to a specific quantum state. Achieving the required level of control to precisely manipulate and measure the desired states without introducing perturbations that could affect the outcome is technically challenging. 
Furthermore, the dynamics of continuously monitored quantum systems pose analytical challenges, emphasising the need for theoretical approaches that can offer valuable insights into the behaviour of these intricate systems. 

\section*{Acknowledgments}
We acknowledge the workshop “Open QMBP 2023” at Institute Pascal (Orsay, France) for hosting us  and as Prof. Ehud Altman for his inspiring lecture on the same workshop, which contributed to the creation of this proceeding. A.S. is grateful for Ermenegildo Zegna Founder's Scholarship for support. A.C. acknowledges support from the EUTOPIA PhD Co-tutelle programme.

\bibliography{biblio.bib}

\end{document}